\documentclass[12pt,a4paper]{article}
\usepackage[T1]{fontenc}
\usepackage[utf8]{inputenc}
\usepackage{authblk}
\usepackage{bbding} 

\usepackage{amsmath}
\usepackage{graphicx,float,color}
\usepackage{hyperref}
\hypersetup{colorlinks = true, linkcolor = blue, urlcolor  = blue, citecolor = blue, anchorcolor = blue}
\usepackage{multirow}
\usepackage[numbers,sort&compress]{natbib}
\usepackage{braket}
\usepackage{amsfonts}
\usepackage{amssymb} 
\usepackage{epic}
\usepackage{eepic}
\usepackage{epsfig}
\usepackage{latexsym}
\usepackage{color}
\usepackage{url}  
\usepackage{caption}
\usepackage[caption=false]{subfig}
\usepackage{float}
\usepackage{bm}
\usepackage{geometry}
\usepackage{slashed}
\usepackage[detect-all]{siunitx}

\def\Mp{m_{\mathrm{Pl}}}
\def\lp{\ell_{\mathrm{Pl}}}

\def\Mmin{M_{\mathrm{min}}}
\def\hMmin{\hat{M}_{\mathrm{min}}}
\def\Mini{M_{\mathrm{init}}}
\def\hMini{\hat{M}_{\mathrm{init}}}

\usepackage{lipsum}
\makeatletter
\def\ps@pprintTitle{%
 \let\@oddhead\@empty
 \let\@evenhead\@empty
 \def\@oddfoot{}%
 \let\@evenfoot\@oddfoot}
\makeatother
\usepackage{etoolbox}

\begin{document}
\title{\textbf{Dark sector production and baryogenesis\\ 
from not quite black holes}}    

\author[]{Ufuk Aydemir\thanks{uaydemir@ihep.ac.cn} }
\author[]{Jing Ren\thanks{renjing@ihep.ac.cn}}
\affil[]{\normalsize Institute of High Energy Physics, Chinese Academy of Sciences, Beijing 100049, P. R. China}

\maketitle

\begin{abstract}
Primordial black holes have been considered as an attractive dark matter candidate, whereas some of the predictions heavily rely on the near-horizon physics that remains to be tested experimentally. 
As a concrete alternative, thermal 2-2-holes closely resemble black holes without event horizons. Being a probable endpoint of gravitational collapse, they not only provide a resolution to the information loss problem, but also naturally give rise to stable remnants. Previously, we have considered primordial 2-2-hole remnants as dark matter. Due to the strong constraints from a novel phenomenon associated with remnant mergers, only small remnants with close to the Planck mass can constitute all of dark matter. 
In this paper, we examine the scenario that the majority of dark matter consists of particles produced by the evaporation of primordial 2-2-holes, whereas the remnant contribution is secondary. The products with light enough mass may contribute to the number of relativistic degrees of freedom in the early universe, which we also calculate.
Moreover, 2-2-hole evaporation can produce particles that are responsible for the baryon asymmetry. We find that baryogenesis through direct B-violating decays or through leptogenesis can both be realized. Overall, the viable parameter space for the Planck remnant case is similar to primordial black holes with Planck remnants. Heavier remnants, on the other hand, lead to different predictions, and the viable parameter space remains large even when the remnant abundance is small.
\end{abstract}

\newpage
{
  \hypersetup{linkcolor=black}
  \tableofcontents
}

\section{Introduction\label{sec:intro}}

Primordial black holes (PBHs)~\cite{1967SvA....10..602Z,Hawking:1971ei,Carr:1974nx,Carr:1975qj,Carr:1976zz} have long been a subject of interest, particularly as a dark matter candidate~\cite{Chapline:1975ojl,GarciaBellido:1996qt,Frampton:2010sw,Belotsky:2014kca,Kawasaki:2016pql}. The abundance of PBHs that survive until now is heavily constrained and only very few narrow windows in the parameter space are still available~\cite{Carr:2016drx,Sasaki:2018dmp,Carr:2020gox,Green:2020jor}. Smaller PBHs could be relevant if  there are remnants left over after the evaporation~\cite{MacGibbon:1987my,Barrow:1992hq,Carr:1994ar,Clark:2016nst,Lehmann:2019zgt,Barrau:2019cuo}.
In fact, provided that the initial PBH mass is small enough that evaporation had been completed before Big Bang Nucleosynthesis (BBN), 
the Planck mass remnants are still a viable dark matter candidate~\cite{Carr:2020gox, Dalianis:2019asr}. 

Alternatively, the contribution of PBHs or their remnants to dark matter could be secondary, while the main component consists of  dark sector particles that have been predominantly produced by PBH evaporation~\cite{Fujita:2014hha, Lennon:2017tqq, Morrison:2018xla, Hooper:2019gtx, Masina:2020xhk,Gondolo:2020uqv}. Since PBHs could reach considerably high temperatures during the evaporation, they can efficiently emit particles in a variety of mass-ranges regardless of the background temperature of the universe. This is in fact relevant in another important context, baryogenesis, especially if the baryon asymmetry is produced in the early universe by heavy particle decays. Additionally, if some of the emitted particles in the dark sector remain relativistic at the time of matter-radiation equality, they could contribute to the radiation content and affect the evolution of the universe.    
  
As one caveat, these discussions heavily rely on the fundamental properties of black holes. However, astrophysical observations only show strong evidences for ultracompact objects that significantly resemble black holes.
Indeed, the Nobel prize for physics in 2020 is given in this context, and there is much more to do observationally to confirm these objects as black holes, regarding in particular the near-horizon physics including the Hawking radiation.\footnote{In fact, it is probably this lack of certainty that made the Nobel committee state in the prize announcement that the prize, in the observational side, is given "for the discovery of a supermassive compact object at the centre of our galaxy", while "black holes" are mentioned in the theoretical part~\cite{nobel2020}.} 
While it is true that General Relativity (GR) is extremely successful in describing the gravitational phenomenon at macroscopic and cosmological scales, it is anticipated to be replaced by a more complete theory of quantum gravity below the Planck scale $\Mp$. Unlike GR, such a theory could accommodate alternatives as dark and compact as black holes, and thus identifying these observations with black holes requires caution.

In fact, there exists such an object called 2-2-hole~\cite{Holdom:2002xy,Holdom:2016nek,Holdom:2019ouz,Ren:2019afg} in quadratic gravity,      a candidate theory of quantum gravity. As a simple extension of GR, by including all possible dimension-four terms, quadratic gravity is renormalizable and asymptotically free at the quantum level~\cite{Stelle:1976gc, Voronov:1984kq, Fradkin:1981iu, Avramidi:1985ki} due to the new massive modes associated with the quadratic curvature terms.\footnote{Quadratic gravity is known to suffer from the ghost problem at the classical level due to the new spin-2 mode. The proposed methods to deal with this pathology mostly involve modifications of the quantum prescription of the theory~\cite{Lee:1969fy, Tomboulis:1977jk, Grinstein:2008bg, Anselmi:2017yux, Donoghue:2018lmc, Bender:2007wu, Salvio:2015gsi, Holdom:2015kbf, Holdom:2016xfn, Salvio:2018crh}. Although there is still no consensus on the resolution of this problem, the theory does provide a more tractable model to visualize near-horizon effects from the high curvature terms.} 
It turns out that the new terms can play a decisive role for ultracompact objects, and the theory predicts 2-2-holes, a new family of solutions absent in GR.  A 2-2-hole is almost as compact as a black hole without horizon. This naturally resolves the information loss paradox, and may leave distinctive imprints in gravitational wave signals that remain to be dedicatedly searched for. In contrast to other candidates, the formation of a 2-2-hole does not rely on exotic forms of matter, so it may serve as the endpoint of gravitational collapse in Nature. 

If black holes turn out to be ultracompact horizonless objects, the relation between PBHs and dark matter physics deserves to be reinvestigated, and 2-2-holes serve as a good example for the study of alternatives.
Since a 2-2-hole has a minimum mass $\Mmin$, a minimal 2-2-hole naturally serves as a stable remnant. A non-minimal  2-2-hole, on the other hand, radiates like a black hole with unusual thermodynamic characteristics, and could produce strong radiation in the early universe. 
In an earlier paper~\cite{Aydemir:2020xfd}, we studied the implications of having the 2-2-hole remnants as dark matter, and derived the observational constraints. We found that the remnant abundance is strongly constrained by a distinctive phenomenon associated with remnant mergers due to evaporation of the merger product, and that only small remnants not much heavier than $\Mp$ can constitute all of dark matter.  

In this paper, we consider the scenario that the remnants are only subdominant at present and the main content of dark matter were produced through primordial 2-2-hole evaporation in the early universe. We investigate the dark sector production and baryon asymmetry generation in this context, by taking into account the observational constraints on the remnant abundance. In particular, we explore the available parameter space with respect to the fundamental parameter $\Mmin$, which not only determines the remnant mass but also appears in the evaporation rate. There is no such feature in the case of PBH with remnants.    

The rest of the paper is organized as follows. The properties of the thermal 2-2-holes are reviewed in Sec.~\ref{sec:22hole}.  The dark sector production is discussed in Sec.~\ref{sec:dark}. The baryon asymmetry generation is studied in Sec.~\ref{sec:baryon}. 
The observational constraints and the implications are discussed in Sec.~\ref{sec:results}. The paper is concluded in Sec.~\ref{sec:summary}.

\section{Preliminaries on thermal 2-2-hole}
\label{sec:22hole}

The action of quadratic gravity includes two additional quadratic curvature terms, the Ricci scalar square and the Weyl tensor square, 
\begin{eqnarray}
\label{action}
S_{\mathrm{QG}}= \frac{1}{16\pi}\int d^4 x \sqrt{-g}\left(m_{\mathrm{Pl}}^2 R-\alpha\; C_{\mu\nu\rho\sigma}C^{\mu\nu\rho\sigma}+\beta R^2\right)\,,
\end{eqnarray}
where $\alpha$ and $\beta$ are dimensionless couplings. These new terms bring in a spin-0 and a spin-2 mode with the tree level masses  $m_0,\,m_2\approx \Mp/\sqrt{\beta},\,\Mp/\sqrt{\alpha}$. As the most generic solution in the theory, the existence of 2-2-holes relies on the Weyl tensor term, and its minimum mass is determined by the mass of the spin-2 mode, $\Mmin\approx \Mp^2/m_2$. In the quantum theory, the dimensionless coupling $\alpha\gtrsim 1$, and hence $\Mmin\gtrsim \Mp$. In the strong coupling scenario, the Planck mass arises dynamically by dimensional transmutation, where $\alpha\approx1$ and $\Mmin\approx \Mp$. In the weak coupling scenario, on the other hand, a large mass hierarchy is allowed and $\Mmin$ could be much larger than the Planck mass.

2-2-holes resemble black holes closely from the exterior, while it features a novel high curvature interior as dominated by quadratic curvature terms.\footnote{The name of the 2-2-holes is related to the special leading order behavior of metric functions around the origin, i.e. $g_{tt}, g_{rr}\propto r^2$, where $r$ is the radial coordinate~\cite{Holdom:2016nek}.} Around the gravitational radius, there is a transition region relating the two distinctive behaviors. For a typical 2-2-hole, the transition region is extremely narrow, and it is hard to tell apart from the current observations. No exotic form of matter is required for the existence of 2-2-holes. As an example,
a thermal gas that is too soft to support an ultracompact configuration in GR is able to source a 2-2-hole in quadratic gravity~\cite{Holdom:2019ouz,Ren:2019afg}. This not only provides a more realistic endpoint for a generic gravitational collapse, but also enables the study for thermodynamics of ultracompact horizonless objects in parallel to the discussion of compact stars in GR~\cite{Sorkin:1981wd}. As a result, the thermodynamic behavior of 2-2-holes is expected to be closely related to the structure of their high-curvature interiors, and this serves as a sharp prediction of the theory. In the following, we first review the thermodynamics and evaporation of 2-2-holes, and then discuss the observational constraints from our earlier work~\cite{Aydemir:2020xfd}.

\subsection{Thermodynamics and evaporation}
Without loss of generality, we focus on 2-2-holes sourced by massless relativistic particles, with the equation of state $\rho=3p$. Following the conservation law of the stress tensor, the local measured temperature satisfies Tolman’s law, and it grows large in the deep gravitational potential in the interior. When the 2-2-hole is not in thermal equilibrium with its surroundings, the temperature at spatial infinity $T$ is the one at which it radiates as a black body. 

Depending on the mass, 2-2-holes could have distinctive thermodynamic behavior.  
For a large 2-2-hole with $M$ considerably larger than $\Mmin$,  the interior thermal gas constitutes a high temperature firewall with a large angular proper length and a rather small radial proper length. As a result, independent of the mysterious features of the event horizon, a large 2-2-hole exhibits anomalous thermodynamics just like black holes, e.g. negative heat capacity and the area law for the entropy. 
A small 2-2-hole with $M$ approaching $\Mmin$ behaves more like a star in GR, with positive heat capacity and the entropy scaling trivially with the interior size. In the minimum mass limit, the temperature at infinity, entropy, and the interior size all approach zero. 
Thus, a large 2-2-hole starts by radiating like a black hole with increasing radiation power. After reaching the maximum temperature at about $1.5\Mmin$, it enters into the remnant stage with negligible radiation. 

The temperature and entropy for a large 2-2-hole can be well approximated as 
\begin{eqnarray}\label{eq:LMlimit}
T\approx 1.7\, \mathcal{N}^{-1/4}\hMmin^{1/2}\, T_\textrm{BH},\quad
S\approx 0.60\, \mathcal{N}^{1/4}\hMmin^{-1/2} \,S_\textrm{BH}\,,
\end{eqnarray}
where $\hMmin\equiv \Mmin/\Mp$, the Hawking temperature $T_\textrm{BH}=\Mp^2/8\pi M$ and the Bekenstein-Hawking entropy  $S_\textrm{BH}=\pi\, r_H^2/\lp^2$. They differ from the black hole quantities only by an overall constant; this introduces additional dependence on  the remnant mass $\Mmin$ and the number of degrees of freedom $\mathcal{N}$ in the thermal gas. Their product remains the same, i.e. $T S = T_\textrm{BH}S_\textrm{BH} = M/2$, in accordance with the the first law of thermodynamics.

A thermal 2-2-hole evaporates when $T$ is larger than the background temperature. Its mass evolution can be described by the Stefan-Boltzmann law, with the power
\begin{equation}\label{eq:SBlaw}
-\frac{dM}{dt}
\approx \frac{\pi^2}{120}\, \mathcal{N}_* \, 4\pi r_H^2 \,T^4\;,
\end{equation}
which assumes $4\pi r_H^2$ as the effective emitted area. $\mathcal{N}_*$ denotes the number of particles lighter than $T$~\cite{MacGibbon:1991tj}, and it could be much smaller than $\mathcal{N}$. 
The time dependences of the temperature and mass take the same form as for a black hole. Treating $\mathcal{N}_*$ as a constant determined by the initial $T$, we have 
\begin{eqnarray}\label{eq:LMlimitTime0}
T(t)\approx T_{\textrm{init}}\left(1-\frac{\Delta t}{\tau_L}\right)^{-1/3},\quad
M(t)\approx \Mini\left(1-\frac{\Delta t}{\tau_L}\right)^{1/3},
\end{eqnarray}
where $\tau_L$ is the evaporation time for a 2-2-hole evolving from a much larger $\Mini$ to $\Mmin$,
\begin{eqnarray}\label{eq:tauL}
\tau_L
\,\approx \,2\times 10^{-40} \, \frac{\mathcal{N}}{\mathcal{N_*}}  \,  \hMmin^{-2} \, \hMini^{3}\; \textrm{s}\,,
\end{eqnarray}
where $\hMini\equiv \Mini/\Mp$. Due to the $\Mmin$ dependence, $\tau_L$ is in general smaller than the lifetime of a black hole with the same mass. 
Note that (\ref{eq:LMlimitTime0}) and (\ref{eq:tauL}) assume evaporation immediately after formation, while primordial 2-2-holes formed in the radiation era may initially have a higher background temperature, and accretion of cosmic radiation needs to be considered. Nonetheless, the growth in the mass is found to be at most of order one and the influence on $\tau_L$ is also negligible~\cite{Masina:2020xhk}. Therefore, we ignore the accretion effects in the following discussion.

For a particle species $j$ with mass $m_j$, by assuming the average energy to be the temperature, the number of particles emitted through the 2-2-hole evaporation is
\begin{eqnarray}
N_j=g_j\int_{t_j}^{\tau_L}\frac{dN}{dt}dt
\approx -g_j\int_{t_j}^{\tau_L}\frac{1}{T}\frac{dM}{dt}dt\,,
\end{eqnarray}
where $g_j$ is the particle species number and $t_j$ denotes the starting time for the emission of particle $j$. Depending on the particle mass, there are two different cases, 
\begin{eqnarray}\label{eq:tchiLH}
\textrm{Light mass case:}&& m_j\leqslant T_\textrm{init},\quad t_j=t_\textrm{init}\,,\\
\textrm{Heavy mass case:}&& m_j> T_\textrm{init},\quad t_j/\tau_L=1-(m_j/T_\textrm{init})^{-3}\,.
\end{eqnarray}
We then find the number of emitted particles 
\begin{eqnarray}  
\label{noofpar}
N_j\approx 7.4\, \kappa_j\,\textrm{B}_j\,  \mathcal{N}^{1/4}\hMmin^{-1/2}\hMini^{2}\,,
\end{eqnarray}
where $\textrm{B}_j=g_j/\mathcal{N}_*$ is the branching fraction and
\begin{eqnarray}
\label{kappa}
\kappa_j=\left\{
   \begin{array}{ll}
   1\,, & \textrm{for}\quad m_j\leqslant T_{\mathrm{init}} \\
   T^2_\textrm{init}/m_j^2\,, & \textrm{for}\quad m_j>T_{\mathrm{init}}
   \end{array}
\right. \,.
\end{eqnarray}
As expected, the number of emitted particles is proportional to the effective emitted area of the hole, and for the heavy particle case it is suppressed by the particle masses. 
Note that we ignore the spin dependence of the number of emitted particles here. Although the effective emitted area in (\ref{eq:SBlaw}) depends on the particle spin in general~\cite{Page:1976df}, given that $T\, r_H\gtrsim1$, the area approaches the geometrical-optics limit regardless of the spin and can be well approximated by the horizon area. Thus, the particle spin plays a little role for our discussion of dark matter production in this paper.

As a final remark, in the Standard Model (SM), the number of particle species varies from 107 to 11 for $T\gtrsim\,$TeV and $T\sim$\,MeV. In most of the expressions, their dependences come with powers smaller than one and hence the choice of different numbers registers errors only on the order of one. Even considering a large dark sector, we restrict to the case that the dark sector contribution is at most at the order of the SM ones. Therefore, for the order-of-magnitude estimation, these factors are simply insignificant. In the rest of the paper, we will suppress the number of species dependence with small powers by using $\mathcal{N}_*=\mathcal{N}\approx107$  and $g_{*}\approx 11$, unless otherwise stated.

\subsection{Observational constraints}

Assuming that the primordial 2-2-holes have already completed the evaporation and become remnants now, the mass fraction of 2-2-hole remnants in dark matter today is
\begin{eqnarray}\label{eq:betaf}
f\equiv\frac{\Mmin\,n(t_0)}{\rho_\textrm{DM}(t_0)}
=\frac{\Mmin\,s(t_0)}{\rho_\textrm{DM}(t_0)}\frac{n(t_0)}{s(t_0)}\,,
\end{eqnarray} 
where $n(t)$ denote the number density for the remnants and $s(t_0)=2.9\times 10^3\,\textrm{cm}^{-3}$, $\rho_{\textrm{DM}}(t_0)\approx 0.26\rho_c$, $\rho_c=9.5\times10^{-30}\,\textrm{g}\,\textrm{cm}^{-3}$~\cite{Aghanim:2018eyx}.\footnote{Another commonly used parameter is the mass fraction at formation $\rho(t_\textrm{init})/\rho_\textrm{tot}(t_\textrm{init})$~\cite{Carr:2009jm}. For 2-2-holes, it is related to the remnant fraction as $\rho(t_\textrm{init})/\rho_\textrm{tot}(t_\textrm{init})\approx 4.0\times 10^{-28} \,f\,\hMmin ^{-1} \,\hMini^{3/2}$, where we have used $T_\textrm{bkg}(t)=0.17 \,\Mp\, (t/\lp)^{-1/2}$ and $\Mini\approx 8\times 10^{37}\left(t_\textrm{init}/\textrm{s}\right)$\,g~\cite{Aydemir:2020xfd}.}

The relation between $f$ and the number density to entropy density ratio at the time of formation $n(t_\textrm{init})/s(t_\textrm{init})$ depends on whether the primordial 2-2-holes have ever come to dominate the energy density or not. Considering the 2-2-hole formation in the radiation era, the initial mass faction of 2-2-holes increases with time usually from a small value. As the leading order approximation for the cosmic evolution, we consider the evaporation as an instantaneous radiation of energy at $t\approx \tau_L$, with the 2-2-hole mass $M(t) \approx \Mini$ at $t \leq \tau_L$ and $M(t) \approx \Mmin$ at $t > \tau_L$.  For a given $\Mini$, we can then define a critical number density at formation,
\begin{eqnarray}
n_c(t_\textrm{init})=\frac{\rho_\textrm{rad}(t_\textrm{init})}{\Mini}\sqrt{\frac{t_\textrm{init}}{\tau_L}}\,,
\end{eqnarray}
with which the 2-2-holes and the radiation have equal energy densities at $t\approx \tau_L$.

When $n(t_\textrm{init})\lesssim n_c(t_\textrm{init})$, the \textit{non-domination} scenario,  the 2-2-holes are always subdominant in the energy budget, and the entropy injection from evaporation is negligible. The ratio $n(t)/s(t)$ remains constant till the present, with $n(t_0)/s(t_0) \approx n(\tau_L)/s(\tau_L)\approx n(t_\textrm{init})/s(t_\textrm{init})$. The mass fraction of remnants today is then,
\begin{eqnarray}\label{eq:betaf}
f\approx 2.6\times 10^{28}\hMmin \frac{n(t_\textrm{init})}{s(t_\textrm{init})}\,.
\end{eqnarray} 
When $n(t_\textrm{init})\gtrsim n_c(t_\textrm{init})$, the \textit{domination} scenario,  2-2-holes become dominant at some earlier time and there is a new era of matter domination before $\tau_L$. It turns out that the extra redshift of the number density introduced by this new era cancels with the large initial density so that $n(\tau_L)$ remains the same as the one with $n_c(t_\textrm{init})$. 
For the thermal radiation, the energy and entropy densities right after evaporation also stay in the same order of magnitude  as the background quantities in the non-domination case, corresponding to the radiation temperature
\begin{eqnarray}\label{eq:TRH}
T_\textrm{bkg}^{\;\tau}\approx 3.4\times 10^{16}\,\hMmin \, \hMini^{-3/2}\;\textrm{GeV}\,,
\end{eqnarray}
at $\tau_L$ for both scenarios.  
Thus, the mass fraction at present has a maximum, 
\begin{eqnarray}
\label{fmax}
f_{\mathrm{max}} 
\approx 2.6\times 10^{28}\hMmin \frac{n_c(t_\textrm{init})}{s(t_\textrm{init})}
\approx 9.4\times 10^{25}\, \hMmin ^{2} \,\hMini^{-5/2}\,,
\end{eqnarray}
and the bound is saturated with $f \approx f_\textrm{max}$ for the domination scenario.

There is a special value of the initial mass $M_\textrm{DM}$ corresponding to $f_\textrm{max}=1$, given as  
\begin{eqnarray}\label{eq:MDM}
M_\textrm{DM}\approx
5.3\times 10^5\,
\hMmin^{4/5}\;\textrm{g}\,.
\end{eqnarray}
Thus, for $\Mini\lesssim M_\textrm{DM}$, with $f_{\textrm{max}}$ being greater than unity, the 2-2-hole remnants can account for all of dark matter, but the 2-2-hole domination is not allowed. For $\Mini\gtrsim M_\textrm{DM}$, even the 2-2-hole domination occurs, the remnants cannot be the majority of dark matter.

For the later discussion of dark matter and baryogenesis, an important input is the 2-2-hole number density to entropy ratio right after evaporation. From (\ref{eq:betaf}) and (\ref{fmax}), we have 
\begin{eqnarray}
\label{novers}
\frac{n(\tau_L)}{s(\tau_L)}=
\left\{\begin{array}{ll}
       3.9\times 10^{-29}\,f\,\hMmin^{-1}\,, &\textrm{non-domination} \\
       3.6\times10^{-3}\, \hMmin\, \hMini^{-5/2}\,, & \textrm{domination}
\end{array}\right. \,.
\end{eqnarray}
The result for the domination case can be found from the non-domination case by setting $f=f_\textrm{max}$.

The evaporation of primordial 2-2-holes are subject to strong constraints from BBN and CMB~\cite{Aydemir:2020xfd}. To evade the bounds, it is safe to have the evaporation ending before BBN, i.e. $\tau_L\lesssim 1\,$s. 
This imposes an upper (lower) bound on the initial mass (temperature),  with
\begin{eqnarray}\label{eq:MBBN}
M_\textrm{BBN}\approx 3.7\times 10^8 \,\hMmin^{2/3}\;\textrm{g} \,,\quad
T_\textrm{BBN}\approx 1.5\times 10^4\,\hMmin^{-1/6}\;\textrm{GeV}\,.
\end{eqnarray} 
As we can see, the special values of the initial mass in (\ref{eq:MDM}) and (\ref{eq:MBBN})  both increases with $\Mmin$, but with different power. Their equality $M_\textrm{DM}=M_\textrm{BBN}$ then defines a special value of the remnant mass,
\begin{eqnarray}
\label{MD}
\Mmin^\textrm{D} \approx 4.7\times 10^{16}\,\textrm{g}\,.
\end{eqnarray}
For small remnants with $\Mmin\lesssim \Mmin^\textrm{D}$, the 2-2-hole domination is allowed for the initial mass range $M_\textrm{DM}\lesssim \Mini\lesssim M_\textrm{BBN}$. For large remnants with $\Mmin\gtrsim \Mmin^\textrm{D}$, we have $\Mini\lesssim M_\textrm{BBN}<M_\textrm{DM}$ and only the non-domination case is relevant.

\begin{figure}[h]
  \centering%
{ \includegraphics[width=11cm]{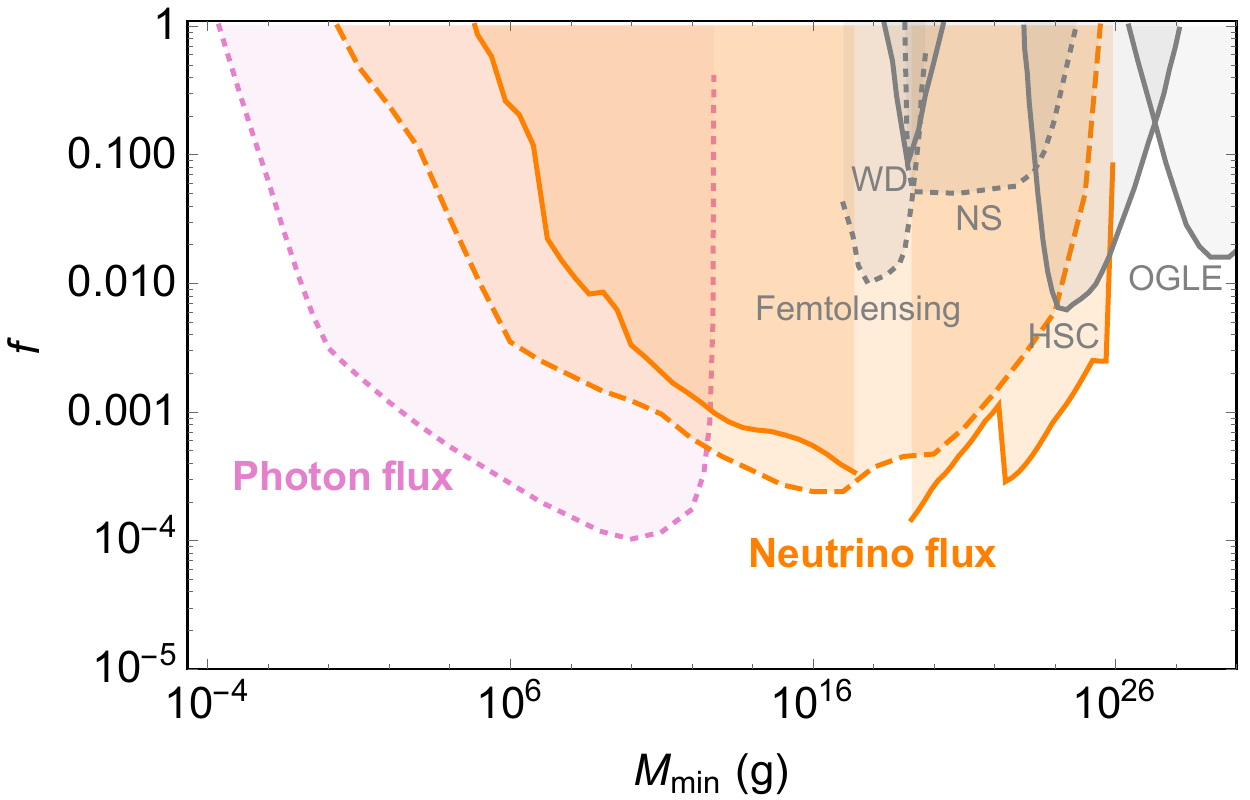}}
\caption{\label{fcons} 
Constraints on the mass fraction of 2-2-hole remnants $f$ as a function of $\Mmin$~\cite{Aydemir:2020xfd}. The gray lines present upper bounds from purely gravitational interactions as in the case for PBHs. The colored lines show the constraints on the high-energy particle fluxes special to 2-2-hole remnants. The solid line considers only the on-shell neutrinos and serves as a conservative estimation. The dash and dotted lines include the parton shower effects and may suffer more from the theoretical uncertainties. } 
\end{figure}

The present observations can directly probe the remnant mass $\Mmin$. Remnants with $\Mmin\gtrsim 10^{17}\,$GeV can be detected through gravitational interaction as in the case of PBHs. Lighter remnants, on the other hand, are  accessible due to a distinctive phenomenon associated with the remnant mergers. 
According to the 2-2-hole thermodynamics, the merger product of remnant binaries with mass around $2\Mmin$ can be quite hot, and its temperature is close to the maximum allowed value, with
\begin{eqnarray}\label{eq:Tmerger}
T_\textrm{merger}\approx 1.3\times 10^{17}\hMmin^{-1/2}\,\textrm{GeV}\,.
\end{eqnarray}  
Thus, the evaporation of the merger product will produce high-energy particle fluxes, with the average energy ranging from the Planck scale down to GeV scale. Considering the latest estimations for the binary merger rate and the parton shower effects for the high-energy emission, we found strong constraints from the photon and neutrino flux measurements for $\Mmin\lesssim 10^{26}\,$g due to this novel phenomenon, as summarized in Fig.~\ref{fcons}. To account for all of dark matter, $\Mmin$ has be to small and the upper bound varies from $10^5\,$g to $10\,\Mp$ depending on the parton show effects.

In the rest of the paper, the following benchmark values of $\Mmin$ are chosen to present the results,
\begin{eqnarray}\label{eq:MminBM}
\Mmin \approx \Mp,\; 10^5\,\textrm{g},\; 10^{28}\,\textrm{g}\,.
\end{eqnarray}  
$\Mmin \approx \Mp$ corresponds to the strong coupling scenario with only one fundamental scale in the theory. $\Mmin\approx 10^5\,$g case has a large uncertainty for the constraints on $f$, which may range from $10^{-4}$ to $1$ depending on whether the parton show effects are included or not. $\Mmin\approx 10^{28}\,$g is around the Earth mass and related to the anomalous microlensing events recently observed by OGLE with $f$ at a per cent level~\cite{Niikura:2019kqi}.\footnote{In order to be consistent with the precise solar-system test of GR, we require the Compton wavelength of the spin-2 mode no larger than $\mathcal{O}$(km). This leads to a rough upper bound $\Mmin\lesssim 10^{33}$\,g that still includes the case with the Earth mass.}


\section{Dark sector production}
\label{sec:dark}

In similarity with the black hole case, the evaporation of primordial 2-2-holes provides a natural production mechanism for the dark sector particles that may only interact with the SM through gravity. In this section, we explore the observational implications for the production of the dark matter and dark radiation.
In Sec.~\ref{sec:DM} for dark matter, we first study the requirement of the observed relic abundance, and then consider the free-streaming constraints for the initially relativistic particles produced by evaporation. 
Light particles that remain relativistic at the time of matter-radiation equality can be considered as dark radiation, and contribute to the effective number of relativistic degrees of freedom $N_{\mathrm{eff}}$. In Sec.~\ref{sec:Neff}, we explore the dark radiation contribution to $N_{\mathrm{eff}}$ and the possible constraints.

\subsection{Particle Dark Matter}
\label{sec:DM}

A large number of proposals have been put forward for dark matter production, including mechanisms such as freeze-out~\cite{Zeldovich1,Zeldovich2,Chiu:1966kg}, freeze-in~\cite{Hall:2009bx}, gravitational production during inflation~\cite{Chung:1998ua,Chung:1998rq,Chung:2001cb}, misalignment mechanism~\cite{Preskill:1982cy,Abbott:1982af,Dine:1982ah,Turner:1989vc}, and production through out-of-equilibrium decays~\cite{Gelmini:2006pq,Gelmini:2006pw}. As in the case of black holes~\cite{Fujita:2014hha, Lennon:2017tqq, Morrison:2018xla, Hooper:2019gtx, Masina:2020xhk,Gondolo:2020uqv}, 2-2-hole evaporation produces particles regardless of the background temperature,  and hence provides a large viable parameter space for the dark matter mass. In order to obtain the strongest relic abundance constraints on the production through 2-2-hole evaporation, we focus on the simplest scenario and ignore contributions from other mechanisms.

The mass fraction of dark matter particle $\chi$ at present is given as 
\begin{eqnarray}
\label{abundance0}
f_{\chi}=\frac{m_{\chi}}{\rho_{\mathrm{DM}}} \frac{n_{\chi}({t_0})}{s(t_0)} s(t_0)\;,
\end{eqnarray}
where $m_\chi$ is the particle mass and $n_{\chi}({t_0})/s(t_0)\approx n_{\chi}({\tau_L})/s(\tau_L)=N_{\chi} n({\tau_L})/s(\tau_L)$. Here, 
$n({\tau_L})/s(\tau_L)$ denotes the number density to entropy ratio for 2-2-holes given in (\ref{novers}). 
$N_{\chi}$ is the number of $\chi$ particles emitted from the evaporation of a single 2-2-hole given in (\ref{noofpar}). Akin to a black hole, depending on whether the particle mass $m_\chi$ is larger or smaller than the 2-2-hole initial temperature, the particle number differs by the factor of $\kappa_\chi$ given in (\ref{kappa}). 
For the light mass case, $m_{\chi}\leqslant T_{\mathrm{init}}$, we obtain
\begin{eqnarray}
\label{OmegaLnum}
f_{\chi}\approx
\left\{\begin{array}{ll}
       2\times 10^{-18} \dfrac{m_{\chi}}{\mathrm{GeV}}\;f \;\textrm{B}_{\chi}\;\hMmin^{-3/2}\;\hMini^{2}\,, &\textrm{non-domination}\vspace{0.2cm} \\
       2\times10^{8}\; \dfrac{m_{\chi}}{\mathrm{GeV}} \;\textrm{B}_{\chi}\; \hMmin^{1/2}\;\hMini^{-1/2}\,, & \textrm{domination}
\end{array}\right. \,.
\end{eqnarray}
For the heavy mass case, $m_{\chi}>T_{\mathrm{init}}$, there is an additional mass suppression in $\kappa_\chi$, and we find 
\begin{eqnarray}
\label{OmegaHnum}
f_{\chi}\approx
\left\{\begin{array}{ll}
       1.3\times10^{17}\;  \left(\dfrac{m_{\chi}}{\mathrm{GeV}}\right)^{-1} \; f\; \textrm{B}_{\chi}\;  \hMmin^{-1/2}\,, &\textrm{non-domination}\vspace{0.2cm} \\
       1.3\times10^{43}\; \left(\dfrac{m_{\chi}}{\mathrm{GeV}}\right)^{-1} \;\textrm{B}_{\chi}\; \hMmin^{3/2}\;\hMini^{-5/2}\,, & \textrm{domination}
\end{array}\right. \,.
\end{eqnarray}
As expected the dark matter abundance $f_\chi$ is proportional to the 2-2-hole remnant abundance $f$. 
The domination case can be arrived from the non-domination case by setting  $f=f_{\mathrm{max}}$ given in (\ref{fmax}), and so $f_\chi$ shows different dependence on the 2-2-hole masses.

\begin{figure}[h!]
\captionsetup[subfigure]{labelformat=empty}
\centering
\hspace{-0.8cm}
\begin{tabular}{ccc}
\subfloat[\quad\quad\quad(a) $(\Mmin, \textrm{B}_\chi)=(m_{\mathrm{Pl}}, 0.01)$]{\includegraphics[width=6.1cm]{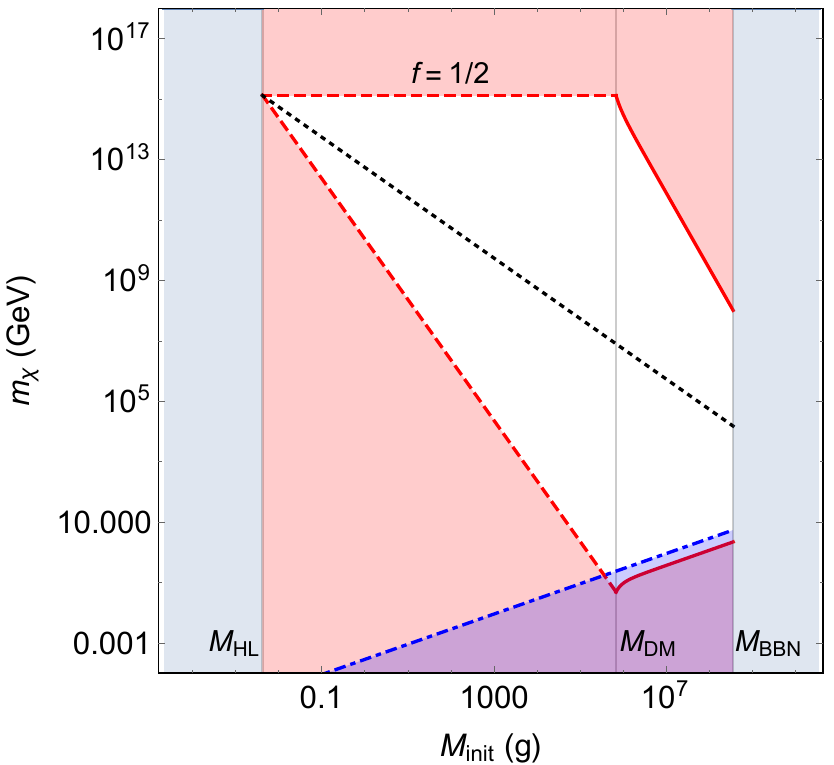}\label{fig:nondom1}}
\subfloat[\quad\quad(b) $(\Mmin, \textrm{B}_\chi)=(10^5\,\textrm{g}, 0.01)$]{\includegraphics[width=5.68cm]{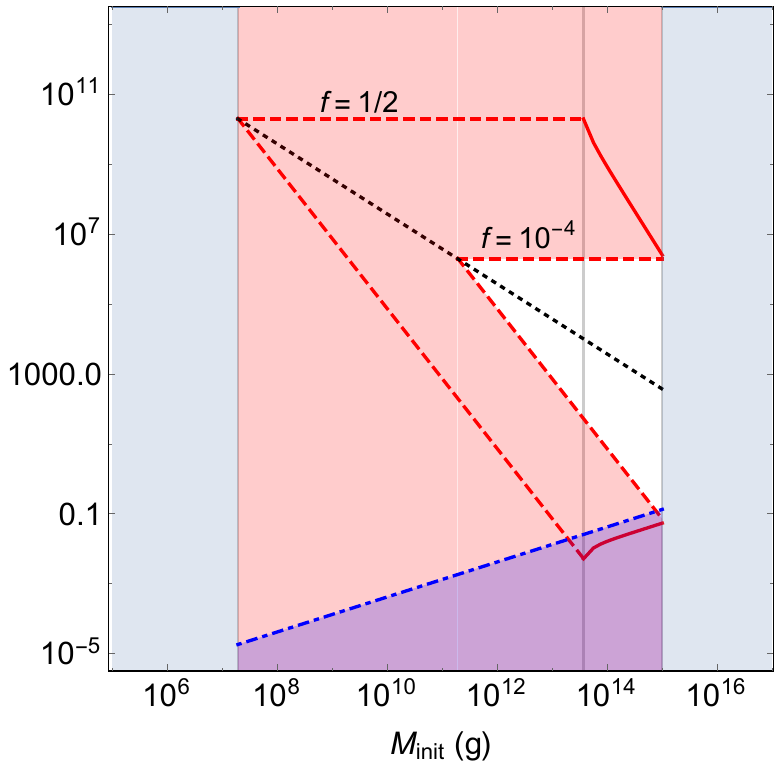}\label{fig:nondom2}}
\subfloat[\quad\quad(c) $(\Mmin, \textrm{B}_\chi)=(10^{28}\,\textrm{g}, 0.01)$]{\includegraphics[width=5.6cm]{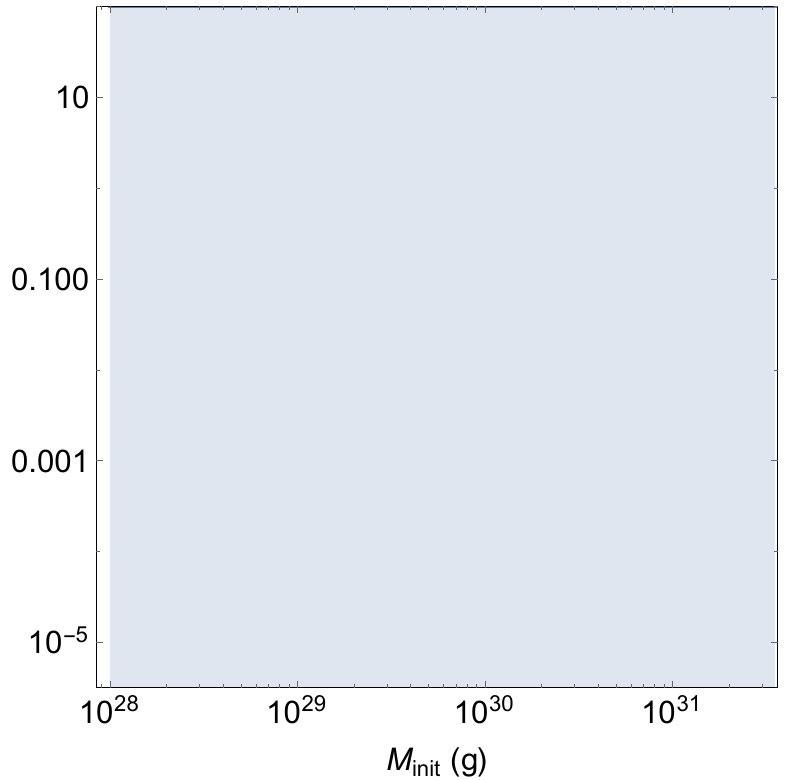}\label{fig:nondom3}}\\
\subfloat[\quad\quad\quad(d) $(\Mmin, \textrm{B}_\chi)=(m_{\mathrm{Pl}}, 0.5)$]{\includegraphics[width=6.1cm]{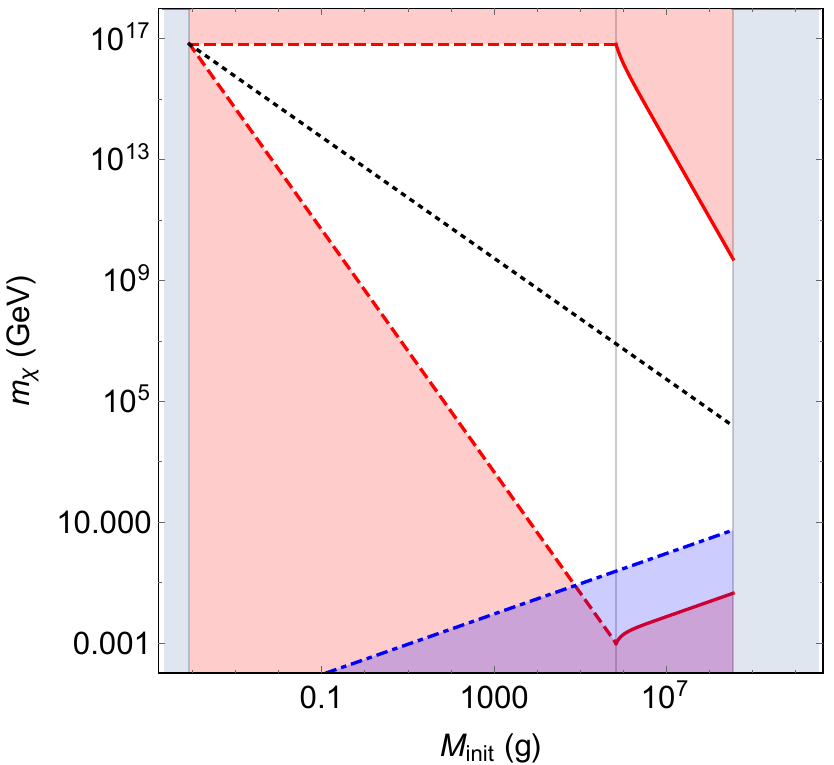}\label{fig:nondom1}}
\subfloat[\quad\quad(e) $(\Mmin, \textrm{B}_\chi)=(10^5\,\textrm{g}, 0.5)$]{\includegraphics[width=5.7cm]{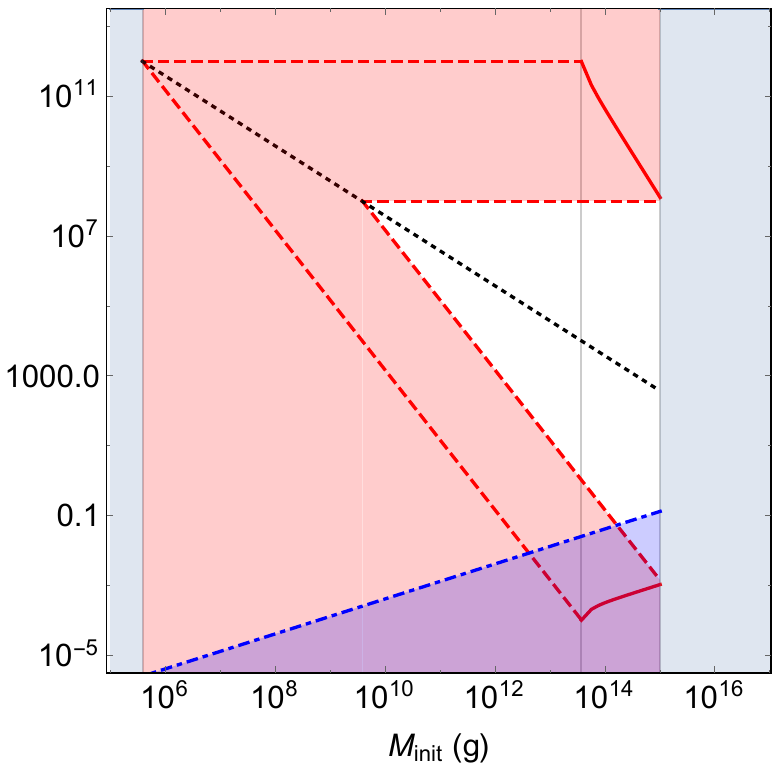}\label{fig:nondom2}}
\subfloat[\quad\quad(f) $(\Mmin, \textrm{B}_\chi)=(10^{28}\,\textrm{g}, 0.5)$]{\includegraphics[width=5.6cm]{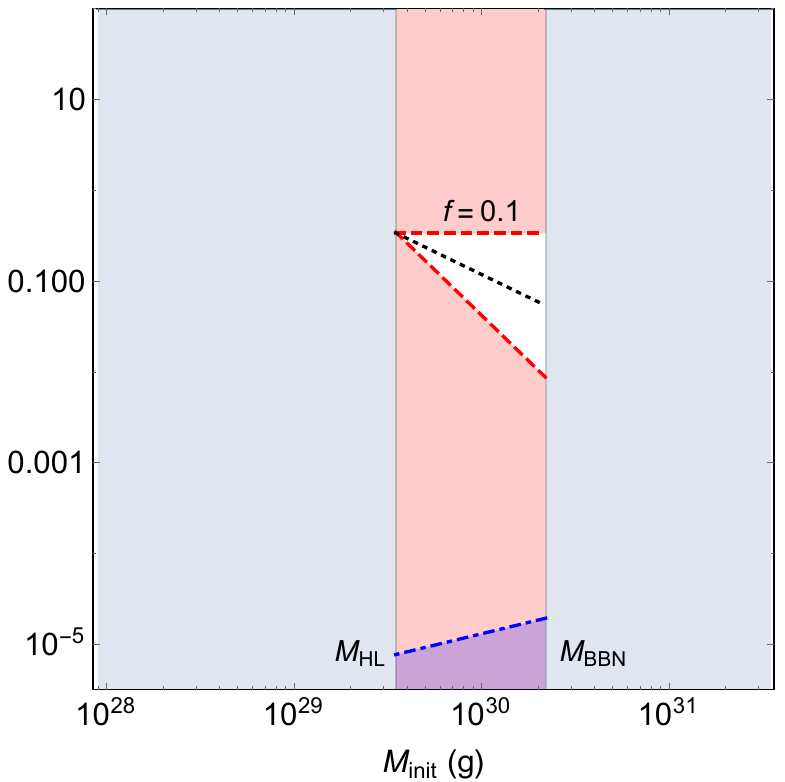}\label{fig:nondom3}}
\end{tabular}
\caption{Constraints on the dark matter mass $m_\chi$ as a function of the 2-2-hole initial mass $\Mini$ for the benchmark remnant masses $\Mmin$ in (\ref{eq:MminBM}), assuming a single particle component with $\textrm{B}_\chi\, (g_\chi)=0.01\,(1)$, $0.5\, (107)$. The white region is allowed, and the black dotted lines denote $T_\textrm{init}$, the separation between the light mass and heavy mass cases. The red dashed lines show the upper and lower bounds derived from the observed abundance in the non-domination case that terminates at $M_\textrm{HL}$ on the left and $M_\textrm{BBN}$ on the right. For small $\Mmin$ in the first and second columns, the 2-2-hole domination is allowed for $M_\textrm{DM}\lesssim \Mini\lesssim M_\textrm{BBN}$, and the thick lines show the relevant parameter space. For the second column, we show the stronger bounds with $f\leq 10^{-4}$ in addition. For the third column, taking $f\leq 0.1$, $M_\textrm{HL}$ goes beyond $M_\textrm{BBN}$ when $\textrm{B}_\chi=0.01$ and there is no viable parameter space. The blue shaded region is excluded by the free-streaming constraints.}
\label{fig:DMabund}
\end{figure}

Figure~\ref{fig:DMabund} presents constraints on the dark matter mass as a function of the initial mass $\Mini$ for some benchmark values of $\Mmin$.
For simplicity,  we assume a single particle component in the dark matter content, in addition to the contribution from 2-2-hole remnants, and hence $f+f_{\chi}=1$.\footnote{In case there are other dark matter production mechanisms in play such as the ones mentioned in the beginning of this subsection, then we would obviously have $f+f_{\chi}\leqslant1$ in order not to overclose the universe. See~\cite{Gondolo:2020uqv} for  discussion of PBHs (without leftover remnants) for the case where there is an additional production mechanism on the top of the production through black hole evaporation.} 
The red dashed lines denote boundaries of the allowed parameter space if the dark matter particle produced by 2-2-hole evaporation account for the observed abundance in the non-domination case. Since $f$ can be arbitrarily small, the abundance constraint only provides an upper and lower bound for $m_\chi> T_\textrm{init}$ (heavy mass) and $m_\chi\leq T_\textrm{init}$ (light mass) respectively. For illustration, we choose $f\leq 1/2$ to show the maximum allowed region, given that we are interested in the case where the particle dark matter is the main component. For $\Mmin=10^5\,$g case, we show the range for $f\leq 10^{-4}$ in addition by considering the observational constraints associated with the remnant mergers. 
We find that the upper bound is independent of $\Mini$, while the lower bound grows large for small $\Mini$. At some small value $M_\textrm{HL}$ the two bounds intersect, and the $\chi$ abundance becomes too small for a smaller $\Mini$ regardless of the dark matter mass. In the white region, there is a one-to-one correspondence between $f$ and $m_\chi$ to satisfy the observed abundance. 
The 2-2-hole domination is allowed for $M_\textrm{DM}\lesssim \Mini\lesssim M_\textrm{BBN}$ when  $\Mmin\lesssim \Mmin^\textrm{D}$ given in (\ref{MD}), and the bounds are saturated with $f_\chi=1-f_\textrm{max}$ (solid lines). For a larger $\Mini$ in this parameter space, the allowed range of $m_\chi$ shrinks due to the decreasing 2-2-holes abundance.

As we can see, the dark matter particles have to be lighter for increasing remnant mass $\Mmin$. 
The upper bound on $m_\chi$ is independent of $\Mini$ and it decreases as $\hMmin^{-1/2}$.  The lower boundary instead is given by the minimum value of the lower dashed lines in Fig.~\ref{fig:DMabund} with $\Mini\approx M_\textrm{DM},\,M_\textrm{BBN}$ for $\Mmin\lesssim \Mmin^\textrm{D}$ and $\Mmin\gtrsim \Mmin^\textrm{D}$, and the remnant mass dependence goes like $\hMmin^{-1/10}$ and $\hMmin^{1/6}$ respectively. Thus, the allowed parameter space shrinks in the weak coupling scenario. For $\Mmin$ as large as the Earth mass $\sim10^{28}\,$g, the allowed mass range is so constrained that the number of degrees of freedom for the dark matter particle can make a big difference.  
For the 2-2-hole domination case, $m_\chi$ cannot stay too close to $T_\textrm{init}$ due to a lower bound on $f_\textrm{max}$ at $\Mini\approx M_\textrm{BBN}$, and the parameter space is more restricted.

Next, we consider the free-streaming constraints. Dark matter particles with too much energy can erase small scale structures and thus they are strongly constrained by observations. In contrast to other mechanisms, particles produced by evaporation are initially relativistic, and only become non-relativistic as the universe expands.
For an order-of-magnitude estimation, we approximate the spectrum by emission at the average energy, and then consider the constraints on the present velocity for the thermal relic~\cite{Fujita:2014hha}. Assuming dark matter particles never reach equilibrium with the thermal bath, the average momentum at present is
\begin{eqnarray}\label{eq:freestreaming1}
p_0=\frac{a(\tau_L)}{a(t_0)}\langle{p(\tau_L)}\rangle\,.
\end{eqnarray}
Up to an order one factor, the average momentum $\langle{p(\tau_L}\rangle\approx T_\textrm{init},\, m_\chi$ for the light mass case ($T_\textrm{init}> m_\chi$) and heavy mass case ($T_\textrm{init}< m_\chi$) respectively. The redshift factor is,
\begin{eqnarray}
\frac{a(\tau_L)}{a(t_0)}\approx\left(\frac{s(\tau_L)}{s(t_0)}\right)^{-1/3}\approx 2.4\times 10^{-30}\,\hMmin^{-1}\,\hMini^{3/2}\,.
\end{eqnarray}
For the dominant component of dark matter, its present velocity $v_0=p_0/m_\chi$ is constrained to be $v_0\lesssim 4.9\times 10^{-7}$~\cite{Viel:2005qj}.

For the light mass case, this imposes a lower bound on the dark matter mass with
\begin{eqnarray}\label{eq:FSLD1}
m_\chi\gtrsim 1.3\times 10^{-6}\,\hMmin^{-1/2}\,\hMini^{1/2}\,\textrm{GeV}\,.
\end{eqnarray}
Due to a smaller amount of the redshift for a larger $\Mini$, the bound increases with $\Mini$. As shown in Fig.~\ref{fig:DMabund}, the free-streaming constraints exclude some part of the parameter space that predicts the observed relic abundance. In particular, the domination scenario for the light mass case is disfavored.  
For the heavy dark matter case, the velocity is independent of $m_\chi$. The maximum value is then found at $\Mini=M_\textrm{BBN}$ independent of other parameters. It turns out to be far smaller than the demanded bound. Thus, there is no constraint for the heavy mass case. As shown in Appendix~\ref{sec:appdA}, these simple estimates are supported by a more informative derivation by considering the momentum distribution and the relativistic fraction of dark matter particles.

\subsection{Dark radiation and the contribution to $N_{\mathbf{eff}}$}
\label{sec:Neff}
A useful way to parameterize the effects of dark radiation is through the change of the effective number of relativistic degrees of freedom $\Delta N_{\mathrm{eff}}$, as defined by 
\begin{eqnarray}
\Delta N_{\mathrm{eff}}=\frac{\rho_{\mathrm{DR}}(t_{\mathrm{EQ}})}{\rho_{\mathrm{R}}(t_{\mathrm{EQ}})}\left[N_{\nu}+\frac{8}{7}\left(\frac{11}{4}\right)^{4/3}\right]\,,
\end{eqnarray}
where $\rho_R(t_\textrm{EQ}), \rho_\textrm{DR}(t_\textrm{EQ})$ are energy densities for thermal radiation and dark radiation at the time of matter-radiation equality, and $N_{\nu}=3.046$ is the standard value for the left-handed neutrinos in the SM~\cite{Mangano:2005cc}.\footnote{A recent analysis by~\cite{Akita:2020szl} estimates a slightly lower value where the difference does not cause a noticeable effect in our analysis.}
A nonzero $\Delta N_{\mathrm{eff}}$ would affect the evolution of the universe, and the current upper limit is $\Delta N_{\mathrm{eff}}\leqslant 0.28$ at 95 \% C.L.~\cite{Bernal:2016gxb}. It may also lead to consequences for the estimation of the Hubble constant. For example, $\Delta N_{\mathrm{eff}}\sim 0.1$ has been suggested~\cite{Bernal:2016gxb,Hooper:2019gtx,Baldes:2020nuv,Dessert:2018khu,Berlin:2018ztp,DEramo:2018vss,Escudero:2019gzq,Shakya:2016oxf} as a resolution for the current Hubble tension~\cite{Riess_2018,Aghanim:2018eyx} between  the local measurements~\cite{Riess:2019cxk,Wong:2019kwg,Freedman:2019jwv} and the CMB-inferred value from Planck~\cite{Aghanim:2018eyx}. Later studies found that changing $\Delta N_{\mathrm{eff}}$ alone is not enough to fully resolve the tension, but the upper limit can be slightly relaxed, $\Delta N_{\mathrm{eff}}\leqslant 0.52$~\cite{Schoneberg:2019wmt}, if the tension is taken into account. In the near future, CMB-S4 measurements might be able to probe $\Delta N_{\mathrm{eff}}\sim 0.02$~\cite{Abazajian:2016yjj}. 

To determine the contribution to $\Delta N_{\mathrm{eff}}$ from 2-2-hole evaporation, we relate the energy densities at the time of matter-radiation equality to the ones right after the end of the 2-2-hole evaporation. In the case of dark radiation, the energy density is simply diluted by the universe expansion, with $\rho_{\mathrm{DR}}(t_{\mathrm{EQ}})a(t_{\mathrm{EQ}})^4=\rho_{\mathrm{DR}}(\tau_L)a(\tau_L)^4$. For the thermal radiation, there are additional contributions from the entropy dumps, and the relation to the scale factor can be found from the entropy conservation $g_{*,\mathrm{eq}}\,a(t_{\mathrm{EQ}})^3 \,T_{\mathrm{EQ}}^3= g_{*,\tau_L}\,a(\tau_L)^3 \,T_{\mathrm{RH}}^3$, where $g_*$ denotes the number of relativistic degrees of freedom at a given time. We then arrive at
\begin{eqnarray}
\frac{\rho_{\mathrm{DR}}(t_{\mathrm{EQ}})}{\rho_{\mathrm{R}}(t_{\mathrm{EQ}})}\approx \frac{\rho_{\mathrm{DR}}(\tau_{L})}{\rho_{\mathrm{R}}(\tau_L)}\frac{g_{*,\mathrm{eq}}^{1/3}}{g_{*,\tau_L}^{1/3}}\;,
\end{eqnarray}
where we have ignored the difference in the definition of $g_{*}$ in entropy and energy since the corresponding error are well within $\mathcal{O}(1)$, hence negligible. $\rho_{\mathrm{R}}(\tau_L)$ can be related to the energy density of 2-2-holes  through the evolution of density ratios, $\rho(\tau_L)/\rho_{\mathrm{R}}(\tau_L)=a(\tau_L)\,\rho(t_{\mathrm{init}})/(a(t_{\mathrm{init}})\,\rho_{\mathrm{R}}(t_{\mathrm{init}}))=f/f_{\mathrm{max}}$, where $f_{\mathrm{max}}$ is given in (\ref{fmax}). In the domination scenario, $f=f_{\mathrm{max}}$ and $\rho(\tau_L)=\rho_{\mathrm{R}}(\tau_L)$.\footnote{Notice that $a(t_{\mathrm{init}})/a(\tau_L)$ corresponds to the critical value for the ratio of initial energy densities such that if it is larger than this value 2-2-holes come to dominate before $t=\tau_L$.} Finally, by using $\rho_{\mathrm{DR}}(\tau_{L})/\rho(\tau_{L})\approx \textrm{B}_{\mathrm{DR}}$, we obtain
\begin{eqnarray}
\label{DeltaNeffgen}
\Delta N_{\mathrm{eff}}\approx
\left\{\begin{array}{ll}
6.6\times 10^{-26}\;\textrm{B}_{\mathrm{DR}}\;f\;g_{*,\tau_L}^{-1/12}\;\hMmin^{-2}\;\hMini^{5/2} \,, &\textrm{non-domination}\vspace{0.2cm} \\
       11.2\;\textrm{B}_{\mathrm{DR}}\; g_{*,\tau_L}^{-1/3} \,, & \textrm{domination}
\end{array}\right. \,,
\end{eqnarray}
where $g_{*,\tau_L}= g_{*,\tau_L}^{\mathrm{SM}}+g_{\mathrm{DR}}$. 
In the domination scenario, since $\rho(\tau_L)=\rho_R(\tau_L)$, $\Delta N_{\mathrm{eff}}$ is identical to the black hole case regardless of whether or not there are remnants left over. Thus, it is easy to achieve $\Delta N_{\mathrm{eff}}\sim 0.1$ for $\mathcal{N}_*\sim 100$, which is still allowed by the current limit. In the non-domination case, there is the additional $\Mmin$ dependence and the 2-2-hole contribution differs from the black hole one in general.

\begin{figure}[h!]
\captionsetup[subfigure]{labelformat=empty}
\hspace{-0.8cm}
\begin{tabular}{rr}
\hspace{-0.5cm}
\subfloat[\quad(a) $\Mmin=m_{\mathrm{Pl}},\,f\leq1$]{\includegraphics[width=6.5cm]{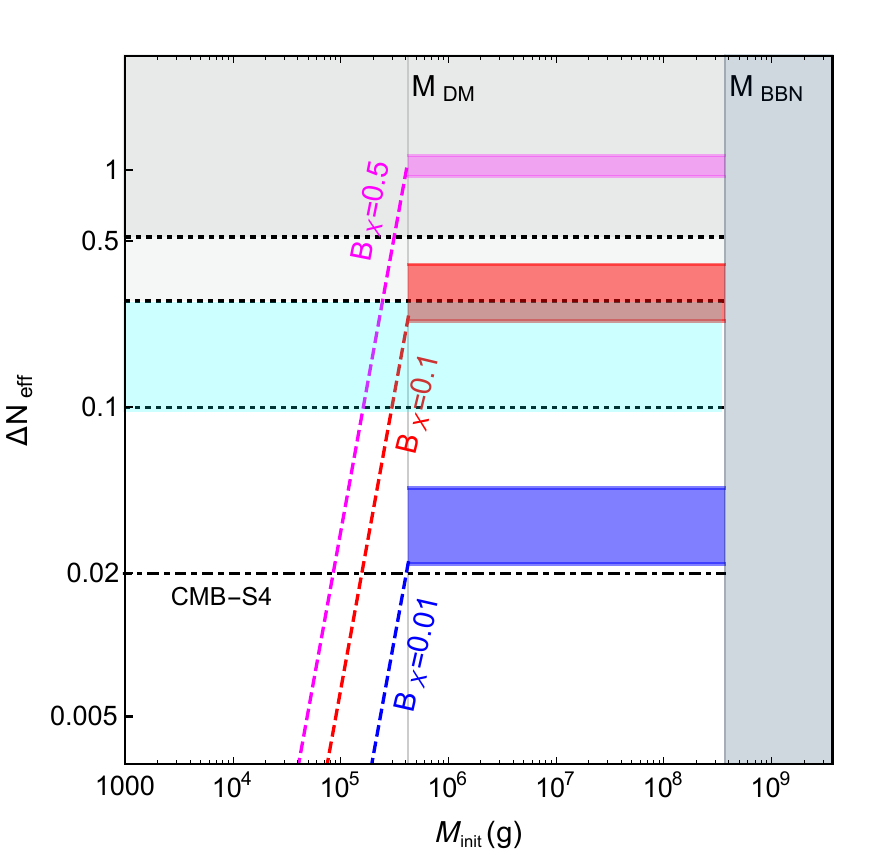}\label{fig:DeltaN1}}\hspace{-0.6cm}
\subfloat[\quad(b) $\Mmin=10^{5}\mathrm{g},\,f\leq10^{-4}$]{\includegraphics[width=6.5cm]{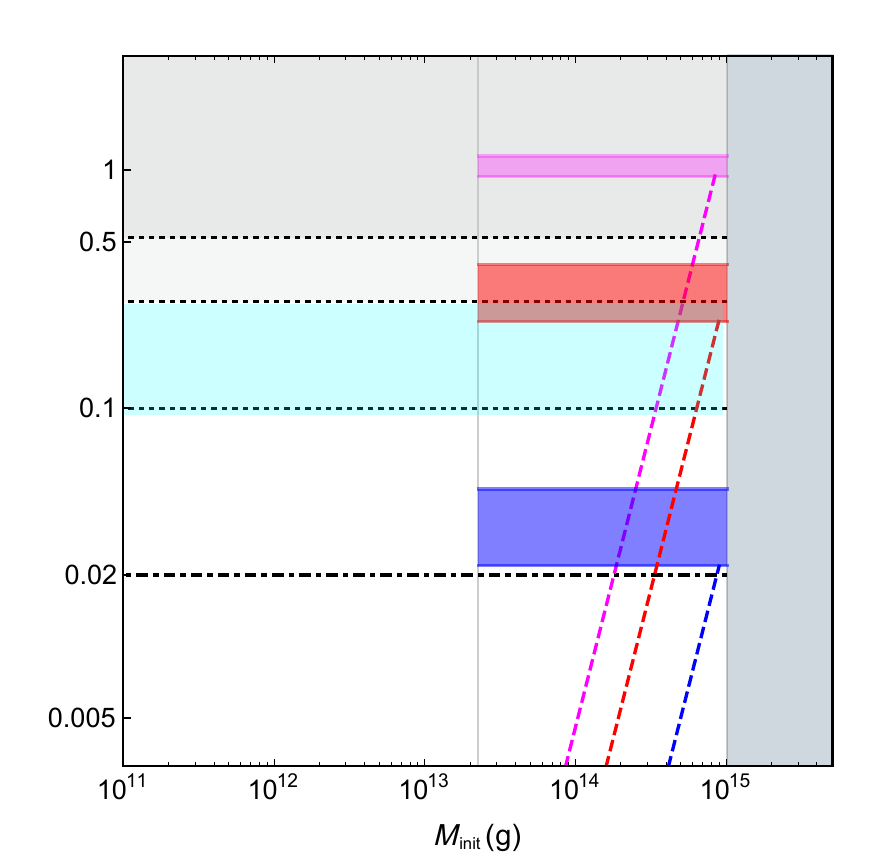}\label{fig:DeltaN2}}\hspace{-0.6cm}
\subfloat[\quad(c)$\Mmin=10^{28}\mathrm{g},\,f\leq 0.1$]{\includegraphics[width=6.4cm]{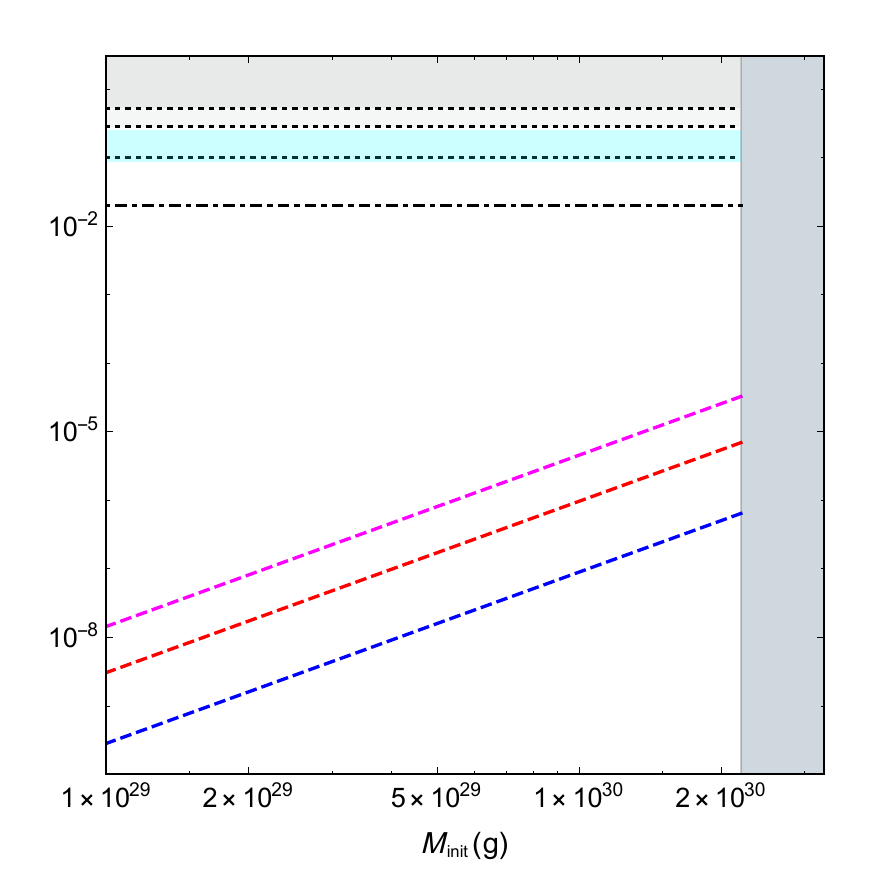}\label{fig:DeltaN3}}
\end{tabular}
\caption{Several benchmark cases demonstrating contributions to $\Delta N_{\mathrm{eff}}$ from the 2-2-hole evaporation in the early universe, for a given $\Mmin$. Blue, red, and magenta lines/bands denotes regions for  $\mathrm{B}_{\mathrm{DR}}$ ($g_{\mathrm{DR}})=0.01\;(1),\; 0.1\;(12)$ and $0.5\; (107)$, respectively. 
The parameter space for the domination scenario, which can be realized only for $ M_{\mathrm{DM}}\lesssim \Mini \lesssim  M_{\mathrm{BBN}}$ when $\Mmin\lesssim  \Mmin^\textrm{D}$,  is given as shaded horizontal bands where the upper and lower limits correspond $g_{*,\tau_L}^{\mathrm{SM}}\approx 11$ and 107, respectively. The non-domination case corresponds to the dashed lines and connects to the domination band at $\Mini$ where $f_{\mathrm{max}}$ saturates the upper bound. 
The grey regions denote the excluded parameter space based on the Planck data, with the upper bound being $\Delta N_{\mathrm{eff}}\leqslant 0.28$ (or $0.52$ if the Hubble tension is taken into account). 
The dotdashed line denotes the projected sensitivity of CMB-S4 measurements. 
The turquoise region shows the parameter space that could potentially alleviate the Hubble tension.}
\label{fig:DeltaN}
\end{figure}

We display the prediction for $\Delta N_{\mathrm{eff}}$ in Fig.~\ref{fig:DeltaN} as a function of $\Mini$ for several $\Mmin$ values. 
For a small $\Mmin$ value for which $f$ is not (significantly) constrained, e.g. Fig.~\ref{fig:DeltaN1}, the maximum contribution is achieved in the domination case at $\Mini\gtrsim M_\textrm{DM}$. Due to the simple form of contribution to $\Delta N_\textrm{eff}$, the measurements can directly constrain the degrees of freedom of dark radiation. The current observations exclude a dark sector with $g_\textrm{DR}\gtrsim 15$ or 35 if the Hubble tension is taken into account, while the future measurements could reach the smallest possible contribution with $g_\textrm{DR}\sim 1$. Below $M_\textrm{DM}$, the contribution to $\Delta N_{\mathrm{eff}}$ in the non-domination case drops quickly with decreasing $\Mini$. For the case that the domination case is excluded, i.e. $g_\textrm{DR}\gg10$, there is only a small range of $\Mini$ that is still currently viable and could be probed in the near future. 
Due to the interplay of $g_{*,\tau_L}$ and $\textrm{B}_\textrm{DR}$ in (\ref{DeltaNeffgen}), increasing $g_{\mathrm{DR}}$ will increase $\Delta N_{\mathrm{eff}}$ up to the range $g_{\mathrm{DR}}\sim 200-300$, above which $\Delta N_{\mathrm{eff}}$ will start to decrease.  
The intersection between the magenta dashed line and CMS-S4 then roughly show the smallest $\Mini$ of interest from the $\Delta N_{\mathrm{eff}}$ measurements.  

For the larger $\Mmin$ cases where $f$ is more strongly constrained, the lower boundary of the domination band shifts to a larger value of $\Mini$ so that $f_{\mathrm{max}}$ satisfies the observational bound. If this value of $\Mini$ is larger than $M_{\mathrm{BBN}}$, then the domination region in the parameter space cannot be reached. For instance, the dashed lines in Fig.~\ref{fig:DeltaN2} barely connect to the domination band. For $\Mmin > \Mmin^{\mathrm{D}}$ given in (\ref{MD}), where the domination scenario cannot be realized, the contribution to $\Delta N_{\mathrm{eff}}$ is in general extremely suppressed. For instance, the prediction in Fig.~\ref{fig:DeltaN3} is still far below the CMB-S4 sensitivity. This situation persists even in the case of a large dark sector with $g_{\mathrm{DR}}\sim\mathcal{O}(100)$.

Finally, in order to consider the evaporation products as relativistic degrees of freedom and include them in $\Delta N_\textrm{eff}$, these particles should have masses smaller than their energies at the time of matter-radiation equality, $\braket{E_{\mathrm{DR}}}|_{\mathrm{EQ}}\approx T_\textrm{init}\, a(\tau_L)/a(t_\textrm{EQ})$. This gives an upper bound on the dark radiation mass, 
\begin{eqnarray}
m_{\mathrm{DR}}\lesssim 0.28\; \hMini^{1/2}\,\hMmin^{-1/2}\;\mathrm{eV}\;.
\end{eqnarray}
For a given $\Mmin$, since this upper bound increases with $\Mini$, the most conservative value can be obtained for $\Mini=M_{\mathrm{BBN}}$, with
\begin{eqnarray}
m_{\mathrm{DR}}\lesssim 1.2\; \hMmin^{-1/6}\;\mathrm{MeV}\;.
\end{eqnarray}
Notice the weak dependence on $\Mmin$. For instance, for $\Mmin=10^{5}\,$g the upper bound becomes $m_{\mathrm{DR}}\lesssim 30\,$keV, and for a much larger value $\Mmin=10^{28\,}$g it is reduced to $m_{\mathrm{DR}}\lesssim 4\,$eV, which is slightly larger than the current limit for the SM neutrino masses.

\section{Baryogenesis}
\label{sec:baryon}

One of the challenges in modern physics is to understand the baryon asymmetry of the universe. Being parameterized by the baryon-to-entropy ratio $\mathcal{B}$, the BBN and CMB observations require  $\mathcal{B}\approx 10^{-10}$~\cite{Zyla:2020zbs}. 
Since an initial contribution can be easily diluted by inflation, the observed baryon asymmetry is usually contemplated to be generated dynamically after reheating. 
It was realized a long time ago by Sakharov that there are three conditions for baryogenesis to occur in the early universe~\cite{Sakharov:1967dj}; existence of baryon number violating interactions, non-conservation of C and CP symmetries, and departure of thermal equilibrium. 
Depending on how these conditions are satisfied, the proposed models for baryogenesis fall in two main categories; out-of-equilibrium decays of heavy particles~\cite{Sakharov:1967dj,Kuzmin:1970nx,Sakharov:1979xt,Toussaint:1978br,Weinberg:1979bt} and electroweak baryogenesis~\cite{Kuzmin:1985mm,Shaposhnikov:1986jp,Shaposhnikov:1987tw,Cohen:1990py,Cohen:1990it,Cohen:1991iu,Nelson:1991ab,Cohen:1993nk}.

The 2-2-hole evaporation can potentially accommodate baryogenesis in both contexts, in similar to black holes (\cite{Barrow:1990he,Baumann:2007yr,Morrison:2018xla,Hooper:2020otu, Zeldovich:1976vw,Dolgov:1980xc,Turner:1979bt,Majumdar:1995yr,Upadhyay:1999vk,Fujita:2014hha,Hook:2014mla,Hamada:2016jnq} and~\cite{Nagatani:1998gv,Aliferis:2014ofa}). For the case of particle decay, the 2-2-hole evaporation could efficiently produce the required heavy particles (accommodated in the theory beyond the SM) regardless of the background temperature.  Moreover, particles emitted by 2-2-holes naturally satisfy the out-of-equilibrium condition as long as they don’t quickly reach thermal equilibrium with the background.\footnote{There are other proposed mechanisms for non-thermal production for particles responsible for baryogenesis, such as production during reheating through the inflaton decay~\cite{Yokoyama:1987hf,SravanKumar:2018tgk}; the Affleck-Dine baryogenesis~\cite{Affleck:1984fy} that utilizes the flat directions of a SUSY potential along which baryon and lepton violation condensates of squarks and sleptons form and subsequently decay to regular fermions;  and its Q-ball version~\cite{Enqvist:1997si,Enqvist:1998en,Fujii:2002kr}.} In contrast in the standard scenario, where the particles are thermally produced, the decay rates are required to fall below the Hubble rate in order for the particles to be out-of-thermal equilibrium with the background. This usually requires the particle to be super-heavy.\footnote{Given the decay rate $\Gamma_X\simeq g_X^2 m_X$ with $g_X$ the coupling and $m_X$ the mass, the out-of-thermal equilibrium condition, $\Gamma_X \lesssim H\sim T_{\mathrm{bkg}}^2 /m_{\mathrm{Pl}}$ at $T_{\mathrm{bkg}}\simeq m_X$, can be satisfied only if $m_X \gtrsim  g_X^2 m_{\mathrm{Pl}}$. For relatively light particles, the thermal production then does not contribute unless they are extremely weakly coupled.} 
As for electroweak baryogengesis, 2-2-hole evaporation above the electroweak scale may satisfy the out-of-equilibrium condition through the domain wall formation outside the hole, without the necessity of a first order phase transition as in the standard scenario.

For both cases, the baryon-to-entropy ratio is given as
\begin{eqnarray}
\label{bartoen}
\mathcal{B}=B\,  \frac{n(\tau_L)}{s(\tau_L)}\,,
\end{eqnarray}
where $B$ denotes the baryon number produced by each evaporating hole. We then obtain 
\begin{eqnarray}\label{eq:BER}
\mathcal{B}\approx
\left\{\begin{array}{ll}
       3.9\times 10^{-29}\,B\,f\,\hMmin^{-1}\,, &\textrm{non-domination} \\
       3.6\times10^{-3}\,B\, \hMmin\, \hMini^{-5/2}\,, & \textrm{domination}
\end{array}\right. \,.
\end{eqnarray}
with the 2-2-hole number density to entropy ratio $n(\tau_L)/s(\tau_L)$ given in (\ref{novers}). Since the remnant abundance is bounded from above by the observed value for dark matter, a heavier remnant is expected to have a smaller number density and then a smaller $\mathcal{B}$. The question is then, what is the mass dependence for the baryon number $B$?   In the following, we discuss the possibility to realize the observed baryon asymmetry in both scenarios, \textit{i.e.} baryogenesis through heavy particle decays and electroweak baryogenesis.

\subsection{Baryon asymmetry from heavy particle decays}

One scenario for the baryon asymmetry generation is through direct baryon number violating decays of heavy particles, which we refer to as "direct baryogenesis". This is generally considered in grand unified theories (GUTs), referred to as GUT baryogenesis~\cite{Harvey:1981yk,Weinberg:1979bt,Nanopoulos:1979gx,Yoshimura:1979gy,Ignatiev:1978uf,Fukugita:2002hu}, which naturally accommodate heavy gauge bosons or colored-scalars that couple to quarks and leptons simultaneously. 
The other scenario, known as leptogenesis~\cite{Fukugita:1986hr,Pilaftsis:1997jf,Hamaguchi:2001gw,Buchmuller:2004nz,Buchmuller:2005eh,Pilaftsis:2005rv,Giudice:2003jh,Davidson:2008bu}, assumes the lepton number  generation through decays of right handed neutrinos first, and then a subsequent conversion to the baryon number  through sphaleron processes. Sphalerons are non-perturbative solutions in the electroweak theory that violate the accidental baryon and lepton numbers conservation at the perturbative level~\cite{Kuzmin:1985mm}. These processes become effective for temperature below $10^{12}\,$GeV and above the electroweak scale.  Sphalerons drive (B+L) to zero, but  they do not effect (B-L). Therefore, any lepton asymmetry at appropriately high energies can be partially converted to baryon asymmetry. In either of these scenarios, if such particles that are responsible for baryogenesis exist in Nature, they would have been emitted  by 2-2-hole evaporation regardless of the underlying theory and their interaction strength with the SM, and the out-of-equilibrium condition would have been easily satisfied. 

\begin{figure}[h!]
\captionsetup[subfigure]{labelformat=empty}
\hspace{-0.8cm}
\begin{tabular}{rr}
\hspace{-0.5cm}
\subfloat[\quad(a) $(\Mmin,\;\mathrm{B}_X)=(m_{\mathrm{Pl}},0.01)$]{\includegraphics[width=6.0cm]{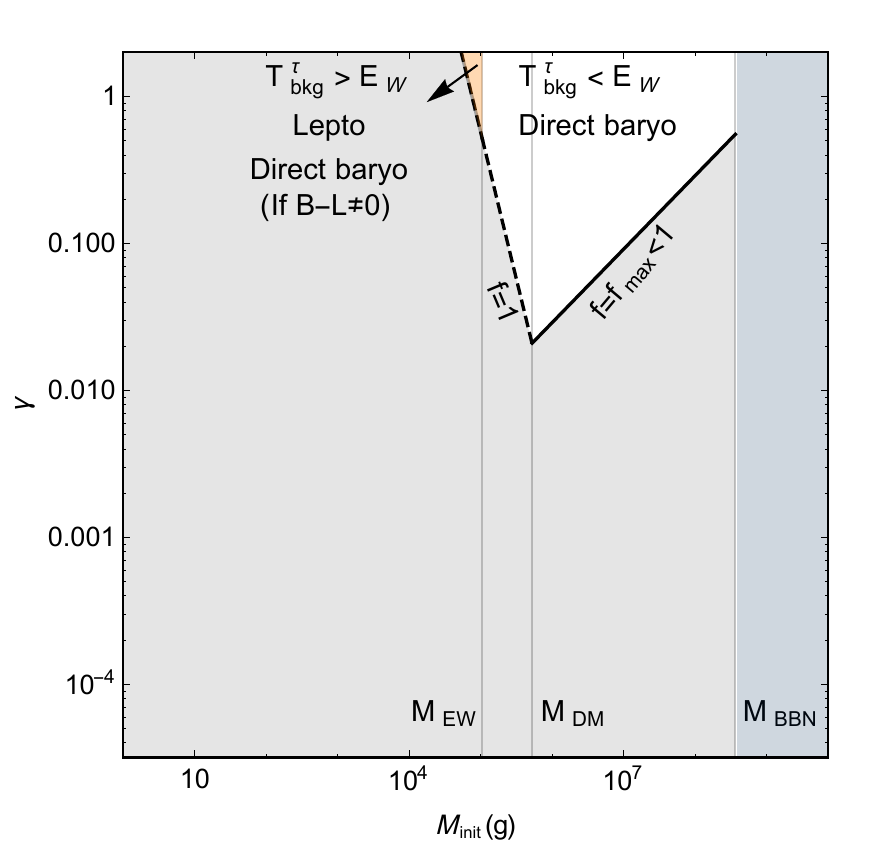}\label{fig:BaryoB1a}}\hspace{-0.6cm}
\subfloat[\quad(b) $(\Mmin,\;\mathrm{B}_X)=(10^{5}\mathrm{g},0.01)$]{\includegraphics[width=6.0cm]{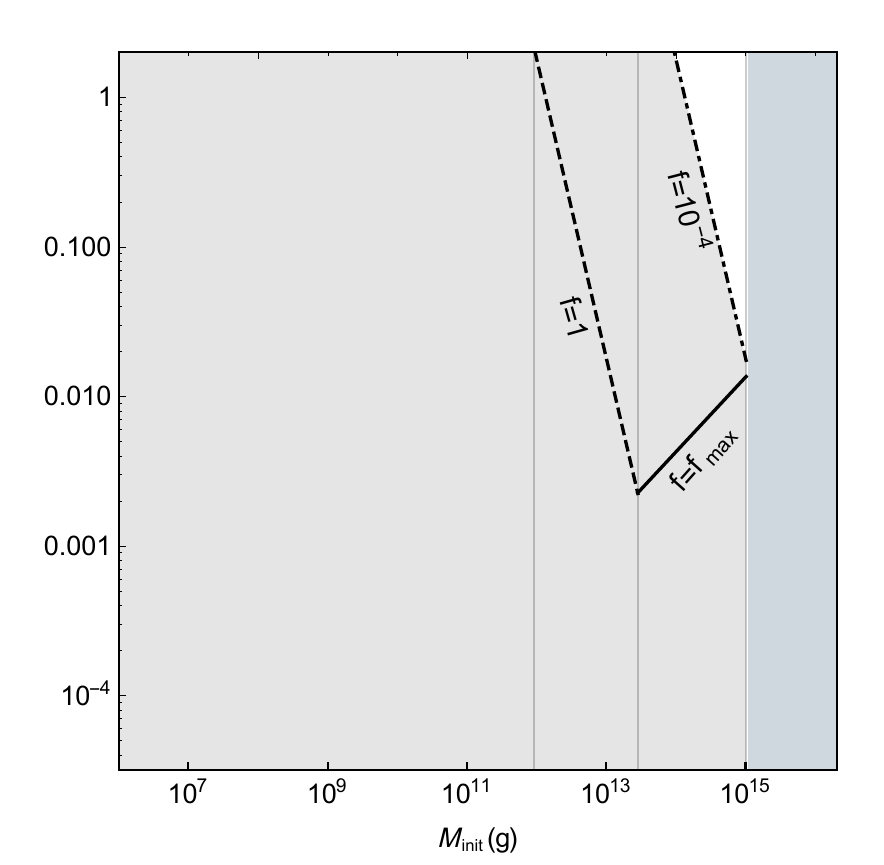}\label{fig:BaryoB2a}}\hspace{-0.6cm}
\subfloat[\quad(c) $(\Mmin,\;\mathrm{B}_X)=(10^{28}\mathrm{g},0.01)$]{\includegraphics[width=6.0cm]{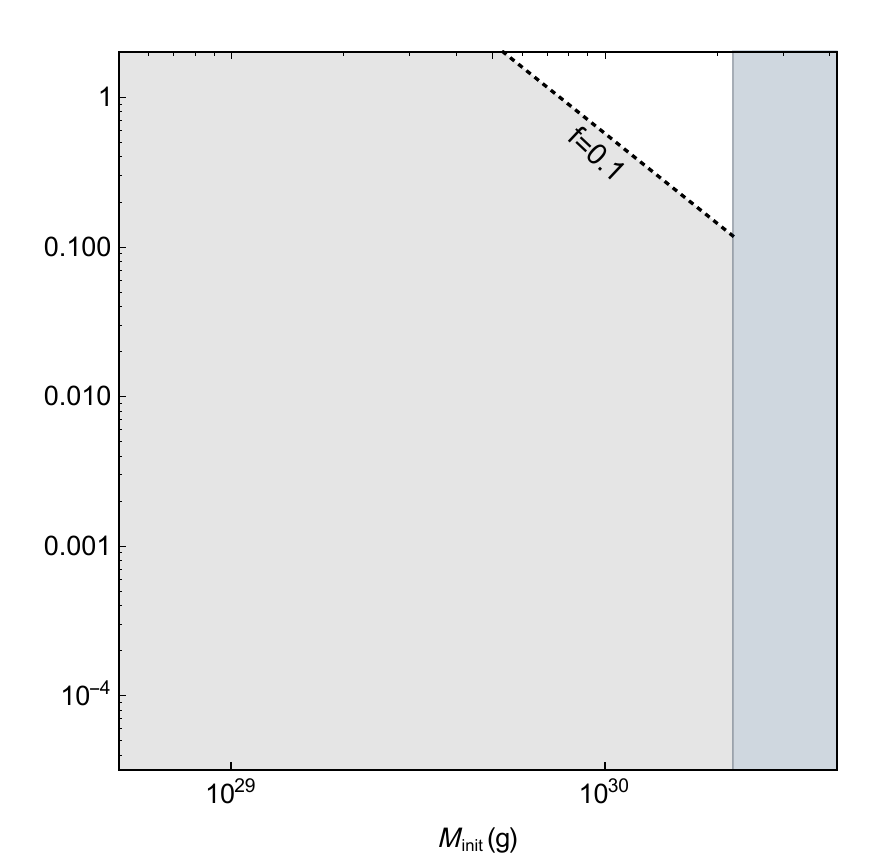}\label{fig:BaryoB3a}}\\
\subfloat[\quad(d) $(\Mmin,\;\mathrm{B}_X)=(m_{\mathrm{Pl}},0.5)$]{\includegraphics[width=6.0cm]{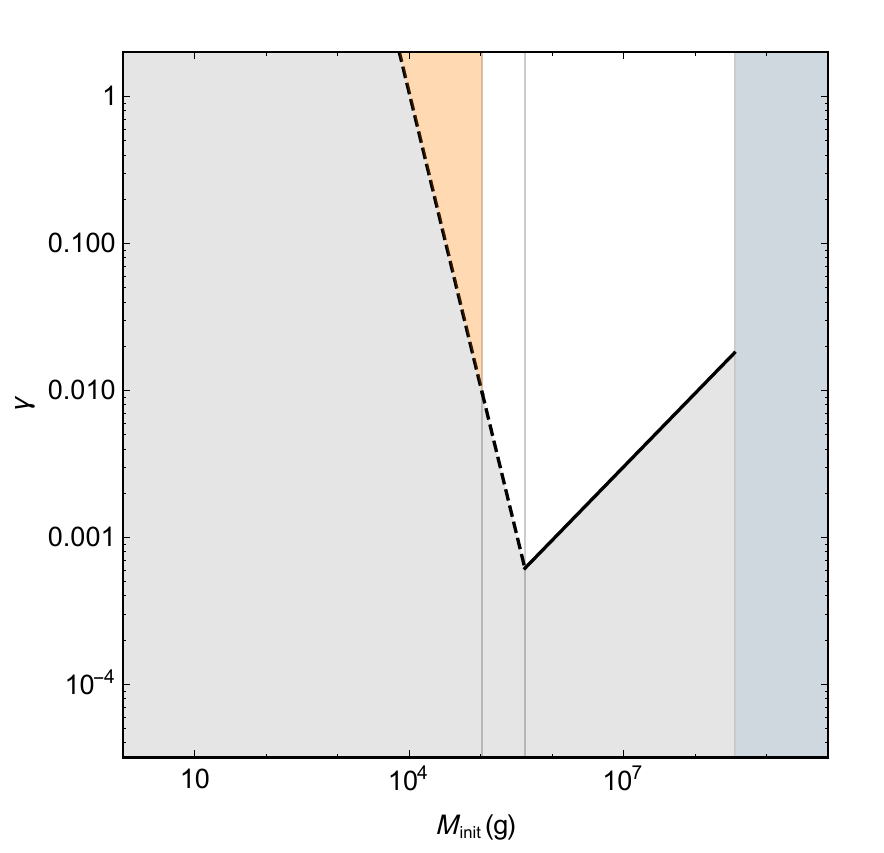}\label{fig:BaryoB1b}}\hspace{-0.6cm}
\subfloat[\quad(e)$(\Mmin,\;\mathrm{B}_X)=(10^{5}\mathrm{g},0.5)$]{\includegraphics[width=6.0cm]{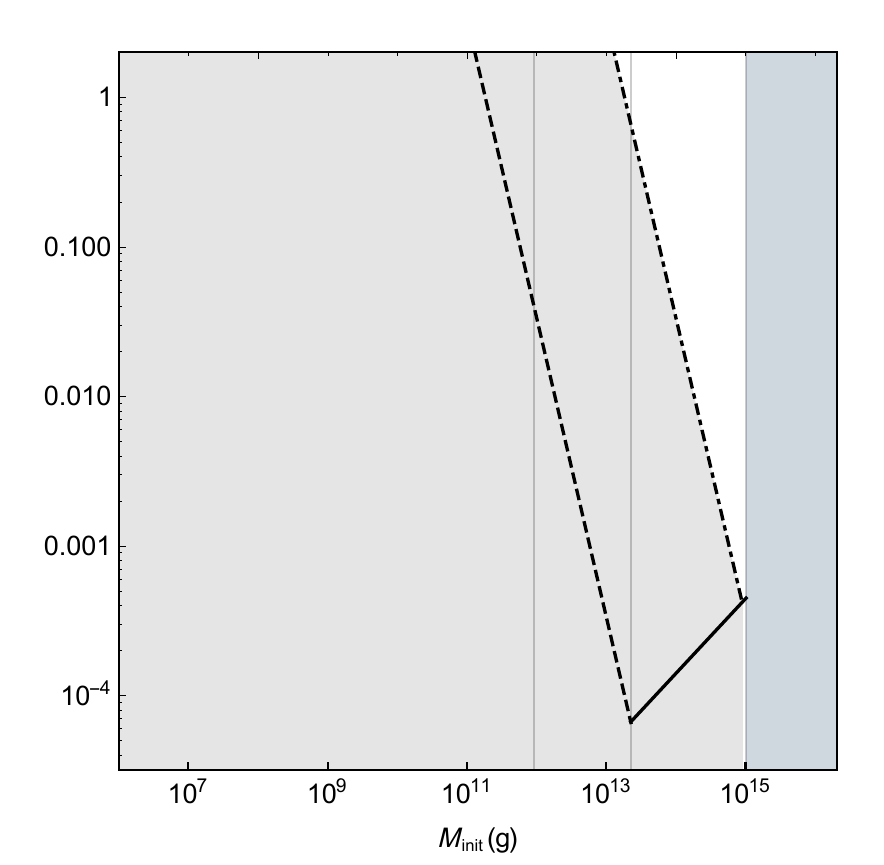}\label{fig:BaryoB2b}}\hspace{-0.6cm}
\subfloat[\quad(f) $(\Mmin,\;\mathrm{B}_X)=(10^{28}\mathrm{g},0.5)$]{\includegraphics[width=6.0cm]{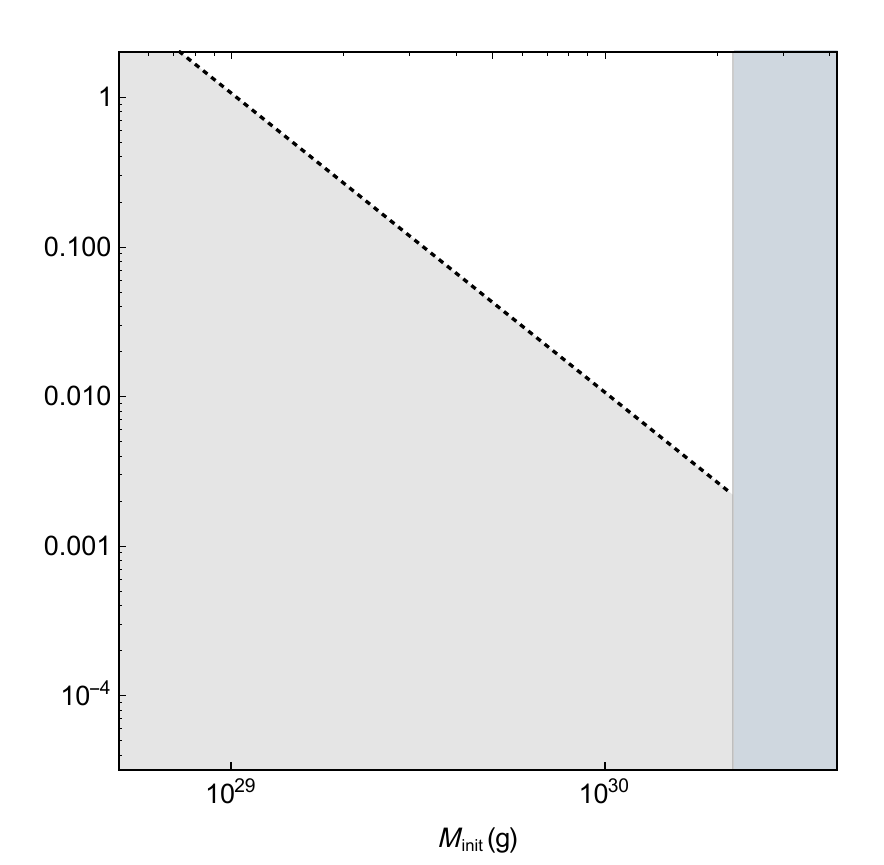}\label{fig:BaryoB3b}}
\end{tabular}
\caption{Constraints on the CP-violation parameter $\gamma$ required for $\mathcal{B} \gtrsim 10^{-10}$, with respect to $\Mini$ for a given $\Mmin$, assuming $\kappa_X=1$ and $\textrm{B}_X\,(g_X)=0.01\,(1)$, $0.5\,(107)$. The grey region denotes the excluded parameter space with $\mathcal{B}$ being too small, and the solid line shows the region relevant for the domination case. For each $\Mmin$ value, we take into account the observational constraints on $f$. $M_\textrm{EW}$ denotes the value of $\Mini$  for which the background temperature $T_\textrm{bkg}^{\;\tau}=E_\textrm{W}$. The orange ($\Mini\lesssim M_\textrm{EW}$) region shows the relevant parameter space for leptogenesis and direct baryogenesis with (B-L) production, while the white ($\Mini\gtrsim M_\textrm{EW}$) region is for direct baryogenesis. 
}
\label{gamma-Minit}
\end{figure}

For both scenarios, assuming the heavy particle $X$ decay promptly,\footnote{The particle decay could be significantly delayed if it is produced with too much kinetic energy and  the scattering with background is too slow to efficiently transfer the energy to background~\cite{Hooper:2020otu}. For 2-2-hole evaporation, the latter happens when $m_X\gtrsim 3\times 10^{16}\,\hMmin^{3/4}\,\hMini^{-9/8}\,$GeV, and this restricts the boost factor $T_\textrm{init}/m_X$ to be less than $380\,\hMmin^{-1/6}$. So the time dilation is not a concern for our order-of-magnitude estimation.} the baryon number produced by each evaporating hole can be written as 
\begin{eqnarray}
\label{bartoen}
B=\gamma\,  N_{X}\approx 24 \,\gamma \, \kappa_X\,\textrm{B}_X\,\hMmin^{-1/2}\,\hMini^{2}\,,
\end{eqnarray}
where the particle number of $X$ is given in (\ref{noofpar}). For direct B-violating decays, $\gamma$ is the parameter that quantifies CP-violation generated through the beyond SM physics and defined as
\begin{eqnarray}
\gamma\equiv \sum_i V_i  \frac{\Gamma (X\rightarrow f_i)-\Gamma (\bar{X}\rightarrow \bar{f}_i)}{\Gamma_X}\;,
\end{eqnarray}
where $V_i$ is the baryon number of the final state $f_i$ and $\Gamma_X$ is the decay width. For leptogenesis, $X$ denotes the right-handed neutrino, $V_i$ becomes the lepton number, and $\gamma$ includes a factor of $\sim 0.65$ due the conversion from leptons to baryons through sphalerons. The value of the parameter $\gamma$ depends on the underlying model and is usually related to the heavy particle mass $m_X$. Different range of values have been predicted in the literature, and $\gamma$ can reach up to  $ \mathcal{O} (1)$, say in resonance leptogenesis~\cite{Pilaftsis:1997jf,Dev:2017wwc}. Here we adopt a model-independent approach, and constrain the parameter space of $\gamma$ in the 2-2-hole evaporation picture.  

 The baryon-to-entropy ratio is obtained from (\ref{eq:BER}) as 
\begin{eqnarray}
\label{B1}
\mathcal{B}=
\left\{\begin{array}{ll}
       9.4\times 10^{-28}\,f\,\gamma \, \kappa_X\,\textrm{B}_X\,\hMmin^{-3/2}\,\hMini^{2}\,, &\textrm{non-domination} \\
       8.8\times 10^{-2}\,\gamma \, \kappa_X\,\textrm{B}_X\,\hMmin^{1/2}\,\hMini^{-1/2}\,, & \textrm{domination}
\end{array}\right. \,,
\end{eqnarray}
where $f=f_{\mathrm{max}}$ is inserted in the domination case. As in the case of black holes~\cite{Hooper:2020otu}, (\ref{B1}) makes implicit assumptions about the background temperature after evaporation, i.e. $T_\textrm{bkg}^{\;\tau}$ in (\ref{eq:TRH}).  If $T_\textrm{bkg}^{\;\tau}$ is larger than the electroweak scale $E_\textrm{W}\approx 100\,$GeV, the sphaleron processes can effectively washout the produced baryon number for direct baryogenesis, and so $T_\textrm{bkg}^{\;\tau}\lesssim E_\textrm{W}$ is required unless there is (B-L) production.\footnote{ Since sphalerons conserve (B-L), a model with a non-vanishing (B-L) number could still provide the baryon asymmetry for $T_\textrm{bkg}^{\;\tau}\gtrsim E_\textrm{W}$. In the GUT context, for instance, this can be realized in the SO(10) theory but not in the SU(5) case~\cite{Riotto:1998bt}.} On the other hand, sphalerons are essential for leptogenesis to transfer the lepton number to baryon number, and then we need $T_\textrm{bkg}^{\;\tau}\gtrsim E_\textrm{W}$ instead. 

Figure~\ref{gamma-Minit} shows the lower limit on $\gamma$ required for $\mathcal{B} \gtrsim10^{-10}$  with respect to $\Mini$ for several benchmark $\Mmin$.\footnote{The main point here is not to get a too small $\mathcal{B}$, which would be completely ruled out by the observations. Overproduction of baryons, on the other hand, can be diluted later in the evolution of the universe~\cite{Baumann:2007yr}.} 
To get the most conservative bound, we take $\kappa_X=1$ to avoid the suppression from the heavy mass $m_X$. We also set $f$ to its maximum allowed value considering observational constraints. For small $\Mmin$ with $f=1$ allowed, i.e. the first column, the minimum required $\gamma$ is achieved at $\Mini\approx M_\textrm{DM}$, the lower boundary of the domination band. In contrast to the black hole case with no remnants~\cite{Baumann:2007yr}, $\gamma$ cannot be further reduced at a smaller $\Mini$ due to the abundance constraints on the 2-2-hole remnants. On the plots, we also highlight a special value of $\Mini$, 
\begin{eqnarray}
M_\textrm{EW}\approx 1.1\times 10^5\,\hMmin^{2/3}\;\textrm{g} \,,
\label{eq:MEW}
\end{eqnarray} 
corresponding to $T_\textrm{bkg}^{\;\tau}=E_\textrm{W}$. Leptogenesis then only operates at $\Mini$ below $M_\textrm{EW}$. Since $M_\textrm{EW}<M_\textrm{DM}$, the minimum $\gamma$ on the plot is relevant for direct baryogenesis, while for leptogenesis a larger $\gamma$, at $\Mini\approx M_\textrm{EW}$, is needed.

For larger $\Mmin$, $f$ is more strongly constrained as in Fig.~\ref{fcons}, and $\gamma$ has to be enhanced to reproduce the observed asymmetry. For leptogenesis, which operates at smaller $\Mini$, the constraint on $\gamma$ is always stronger for larger $\Mmin$. For direct baryogenesis, although the smallest required value of $\gamma$ decreases mildly for larger $\Mmin$ with unconstrained $f$, the final result is sensitive to the exact upper bound on $f$. For instance, for $f=1$, the bound at $\Mini\approx M_\textrm{DM}$ in Fig.~\ref{fig:BaryoB2a} is smaller than that in Fig.~\ref{fig:BaryoB1a}. However, since  $f$ is actually much smaller, the lowest point is lifted to a larger value at $\Mini$ around $M_\textrm{BBN}$. 
For $\Mmin\gtrsim 2.8\times 10^{24}\,$g, $T_\textrm{bkg}^{\;\tau}$ is below $E_\textrm{W}$ for all possible $\Mini$ and leptogenesis becomes irrelevant. For such cases, the 2-2-hole initial temperature is also low. Thus, although the minimum required $\gamma$ could be small, say for $\Mmin\approx 10^{28\,}$g, a quite small $m_X$ is needed as well.

\begin{figure}[h!]
  \centering%
{ \includegraphics[width=11cm]{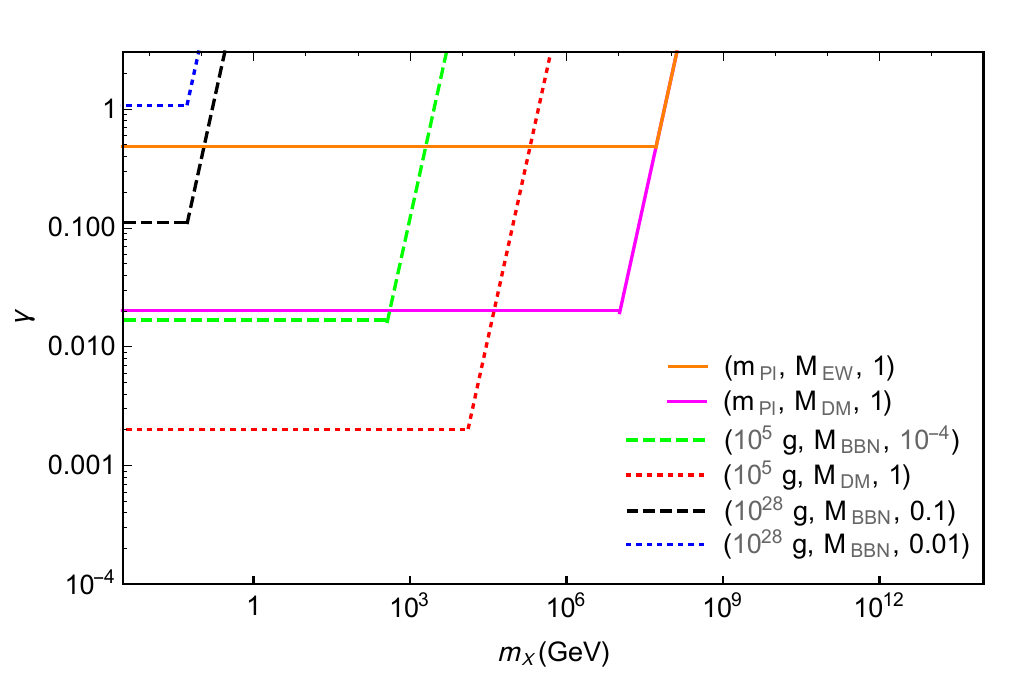}}
\caption{\label{MXCP} 
Constraints on the CP-violation parameter $\gamma$ for $\mathcal{B} \gtrsim 10^{-10}$ with respect to the heavy particle mass $m_X$, assuming $\mathrm{B}_X=0.01$. Benchmark values of $(\Mmin,\;\Mini,\;f)$ are chosen according to Fig.~\ref{gamma-Minit}. For each contour, the horizontal part denotes the light mass case with $m_X \leqslant T_{\mathrm{init}}$ and $\kappa_X =1$, and the ascending part is for the heavy mass case with $m_X > T_{\mathrm{init}}$ and $\kappa_X <1$.  The available parameter space is the upper left region. Only the region corresponding to the orange contour allows leptogenesis. }
\end{figure}

To see the dependence on heavy particle mass $m_X$ more clearly, we present the constraints on the $m_X-\gamma$ plane in Fig.~\ref{MXCP}. The benchmark values chosen to draw the contours correspond to several critical points in Fig.~\ref{gamma-Minit}, for which the limit coincides with the horizontal part for the light mass case. 
The breaking point corresponds to $m_X=T_{\mathrm{init}}$, beyond which the lower limit of $\gamma$ climbs up sharply to compensate the suppression in the total number of particles emitted due to $\kappa_X<1$. Hence, $m_X$ too heavier than the initial 2-2-hole temperature quickly becomes disfavoured given that $\gamma\lesssim \mathcal{O}(1)$. It turns out that the lower limit of $\gamma$ for the heavy mass case is independent of $\Mini$. Different choices of $\Mini$ then only change the breaking point for $m_X$ and the minimum allowed value of $\gamma$. For instance, the orange and magenta solid lines show the most conservative constraints for leptogenesis and direct baryogengesis, respectively, for the Planck remnant case. Due to the smaller allowed value of $\Mini$, leptogenesis is subject to a stronger bound on $\gamma$, but with a larger breaking point for $m_X$. 
The mass $m_X$ is more confined towards smaller values for the heavier remnant case. For instance, in the $\Mmin=10^{5}$\,g case shown by the green line the upper limit becomes the electroweak scale, and for larger $\Mmin$ it decreases further forcing the decay particles to be very light.


\subsection{Electroweak baryogengesis}

Electroweak baryogengesis (EWBG)~\cite{Kuzmin:1985mm,Shaposhnikov:1986jp,Shaposhnikov:1987tw,Cohen:1990py,Cohen:1990it,Cohen:1991iu,Nelson:1991ab,Cohen:1993nk} has been an attractive scenario since it utilizes the sphaleron process in the SM for the baryon number violation, while new physics around the TeV scale is expected to satisfy the other two Sakharov conditions. In the standard scenario, the generation of baryon asymmetry proceeds through bubble nucleation during the electroweak weak phase transition. For a successful baryogenesis, new CP-violation source is required in addition to the one provided by the CKM matrix. The out-of-equilibrium condition can be realized if the electroweak phase transition is strongly first-order to prevent the wash-out of the produced baryon asymmetry. This usually indicates the modification of the Higgs potential, and a large deviation of the Higgs self-interactions, which serve as an important target for the future collider. New physics models that incorporate both ingredients have been extensively studied to realize baryogenesis~\cite{Riotto:1998bt,Morrissey:2012db,Shaposhnikov:2009zzb}. 

As an alternative, it was argued in~\cite{Nagatani:1998gv} that primordial black holes can play a role in EWBG.  If the Hawking radiation temperature is above the electroweak scale, the region surrounding the black hole is the electroweak symmetric phase, and a domain wall separating the symmetric phase from the broken one can form at some large radius. With the sphaleron process taking place in the domain wall near the symmetric region, a sufficient amount of the baryon asymmetry can be generated without the need of a first order phase transition since the Hawking radiation is already a non-equilibrium process. The additional CP violation should still be provided with new physics at around the TeV scale. A close to the maximal CP violation is required for the simplest new physics scenario~\cite{Nagatani:1998gv}, while more involved models could possibly produce  sufficient CP violation~\cite{Aliferis:2014ofa,Aliferis:2020dxr}. In this subsection, we adopt the approach of~\cite{Nagatani:1998gv} to investigate if the situation could be improved for the 2-2-hole evaporation due to the additional remnant mass $\Mmin$ dependence.

Assuming that the evaporation temperature $T(t)$ is much larger than the electroweak scale $E_\textrm{W}$, the emitted particles can reach local thermal equilibrium at some radius larger than the mean-free-path, and from the transfer energy equation, the temperature profile takes the form
\begin{eqnarray}
T(t, r)\approx \left(T_{\text{bkg}}^3(t)+T_0^3(t) \, \frac{r_0}{r}\right)^{1/3} \approx \left(T_{\text{bkg}}^3(t)+1.3\times 10^{-4}\, \hMmin \, \frac{T(t)^2}{r}\right)^{1/3}\,.
\end{eqnarray}
$T_0(t)$ is on the  order of $T(t)$ in (\ref{eq:LMlimitTime0}) and related to the boundary condition close to the would-be horizon, and it is typically much larger than the background temperature. By assuming that the total out-going energy flux equals the Hawking radiation flux, the boundary condition can be fixed as in the last expression. 
If the electroweak phase transition is the second order, a domain wall forms at the radius $r>r_\textrm{DW}$, with
\begin{eqnarray}\label{eq:rDW}
r_\textrm{DW}\approx  1.3\times 10^{-4} \, \hMmin \,\frac{T(t)^2}{E_\textrm{W}^3}\,
\end{eqnarray}
given by the condition $T(t, r_\textrm{DW})\approx E_\textrm{W}$. 
The Higgs vacuum expectation value turns nonzero at $r_\textrm{DW}$ and saturates the broken phase value at $r_\textrm{DW}+d_\textrm{DW}$, and thus $d_\textrm{DW}\approx r_\textrm{DW}$ defines the width of the domain wall.
The mean velocity of the out-going diffusing particles at the domain wall is
\begin{eqnarray}\label{eq:vDW}
v_\textrm{DW}\approx \frac{10}{3}\frac{T_0^3 r_0}{r_\textrm{DW}^2 E_\textrm{W}^4}
\approx  2.6\times 10^4 \, \hMmin^{-1} \,\frac{E_\textrm{W}^2}{T(t)^2}\,.
\end{eqnarray}
Thus, for a heavier remnant, the domain wall grows larger and the particles diffuse slower. 

The emitted particles passing through the domain wall can acquire a nonzero baryon asymmetry by the sphaleron process. With the domain wall properties given in (\ref{eq:rDW}) and (\ref{eq:vDW}), the production rate of the baryon number is
\begin{eqnarray}
\dot{B}\approx 120\pi \, \alpha_W^5\, E_\textrm{W}^3\, r_\textrm{DW}^2\, v_\textrm{DW} \,\epsilon \,\Delta \theta
\approx 6.1\times 10^{-11}\Delta \theta\, \hMmin\,\frac{T(t)^2}{E_\textrm{W}}\,,
\end{eqnarray}
where $\alpha_W\approx g^2/4\pi$ and $\epsilon\approx 1/100$. $\Delta \theta$ is the CP phase, with the typical value $\Delta \theta \sim\pi$.  Integrating the production rate over time, the total baryon number produced during the 2-2-hole evaporation is
\begin{eqnarray}\label{bartoen2}
B\approx \int^{\tau_L}_{t_\textrm{init}}\dot{B}\,dt
\approx (3\tau_L)\left(6.1\times 10^{-11}\,\Delta \theta\, \hMmin\,\frac{T_\textrm{init}^2}{E_\textrm{W}}\right)
\approx 3.8\times 10^7\,\Delta \theta\, \hMini\,.
\end{eqnarray}
The $\Mmin$ dependences in the evaporation time $\tau_L$ and in the rate $\dot{B}$ cancel, and the total asymmetry $B$ only depends on the initial mass. The validity of this derivation assumes two conditions: the size of the domain wall $d_\textrm{DW}$ is greater than the mean-free-path $\sim 10/E_\textrm{W}$ and the evaporation time $\tau_L$ is much larger than the construction time of the domain wall $\sim r_\textrm{DW}/v_\textrm{DW}$. These in turn restrict $\Mini$ within the following range,
\begin{eqnarray}\label{eq:cond}
7.5\times 10^{4}\,\hMmin^{6/7}\,\textrm{g} \lesssim \Mini\lesssim 2\times 10^8\,\hMmin\,\textrm{g}\,.
\end{eqnarray}
The lower bound becomes incompatible with the BBN constraints $\Mini\lesssim M_\textrm{BBN}$ for a too heavy remnant, and EWBG is relevant only for $\Mmin\lesssim 5.5\times 10^{14}\,$g.

From (\ref{eq:BER}), we find the baryon-to-entropy ratio to be
\begin{eqnarray}
\label{B2}
\mathcal{B}=
\left\{\begin{array}{ll}
       1.5\times 10^{-21}\,f\,\Delta \theta \,\hMmin^{-1}\,\hMini\,, &\textrm{non-domination} \\
       1.3\times 10^{5}\,\Delta \theta \,\hMmin\,\hMini^{-3/2}\,, & \textrm{domination}
\end{array}\right. \,.
\end{eqnarray}
Due to the restriction on $\Mmin$ from (\ref{eq:cond}), the maximum allowed value $\mathcal{B}_\textrm{max}\approx 3.7\times 10^{-11}\,\Delta\theta\,\hMmin^{-1/5}$ when $f=1$ and $\Mini=M_\textrm{DM}$, and it decreases for heavier remnants even without considering the stronger constraint on $f$. This can already be seen from the total baryon number in (\ref{bartoen2}). Comparing with (\ref{bartoen}) for the production through heavy particle decay, it receives smaller enhancement from the 2-2-hole mass, and this is not enough to compensate the decrease in the 2-2-hole number density for a large $\Mmin$.
The Planck remnant case, as in the case of PBHs, can barely achieve the observed value $\mathcal{B}\approx 10^{-10}$ with a quite large CP-violating phase $\Delta \theta\approx \pi$. 
Thus, given that $\Mmin\gtrsim \Mp$, dependence on the additional mass scale $\Mmin$ doesn't improve the situation and EWBG is disfavored in the context of 2-2-hole evaporation.


\section{Discussion}
\label{sec:results}
As a concrete example for horizonless ultracompact objects, thermal 2-2-holes not only mimic black holes from many aspects, but also make distinctive predictions for the observations. 
In this paper, we explore the dark sector production and baryon asymmetry generation through the evaporation of primordial thermal 2-2-holes. 
Unlike in the case of a black hole,  a cold remnant is left behind at the end of the 2-2-hole evaporation. The remnant mass $\Mmin$ is determined by the interaction strength in quadratic gravity, with $\Mmin\approx\Mp$ for the strong coupling scenario and $\Mmin\gg \Mp$ for the weak coupling scenario. The same mass parameter also influences the temperature, and makes the 2-2-hole evaporation quantitatively different from a black hole counterpart. The initial mass $\Mini$ is constrained by observations. In order not to contradict the abundance of light elements, 2-2-holes have to evaporate prior to BBN and $\Mini\lesssim M_\textrm{BBN}$ as given in (\ref{eq:MBBN}).
For small remnant with $\Mmin\lesssim \Mmin^D \approx 4.7\times 10^{16}\,$g, a 2-2-hole domination era at early universe is allowed as in the case of black holes, for $\Mini\gtrsim M_\textrm{DM}$ as given in (\ref{eq:MDM}). For a larger $\Mmin$, a 2-2-hole domination era leads to a too large remnant abundance $f$ that will overclose the universe for any $\Mini\lesssim M_\textrm{BBN}$, and is thus forbidden.  
In comparison to PBH cases, the production through 2-2-hole evaporation predicts quite similar parameter space for the strong coupling scenario. For the weak coupling scenario, $f$ is generally more strongly constrained, and different regions of parameter space are inferred.

\subsection{Dark matter and dark radiation}

\begin{figure}[h!]
\captionsetup[subfigure]{labelformat=empty}
\centering
\hspace{-0.8cm}
\begin{tabular}{lll}
\subfloat[(a)\quad With unconstrained $f$]{\includegraphics[width=8cm]{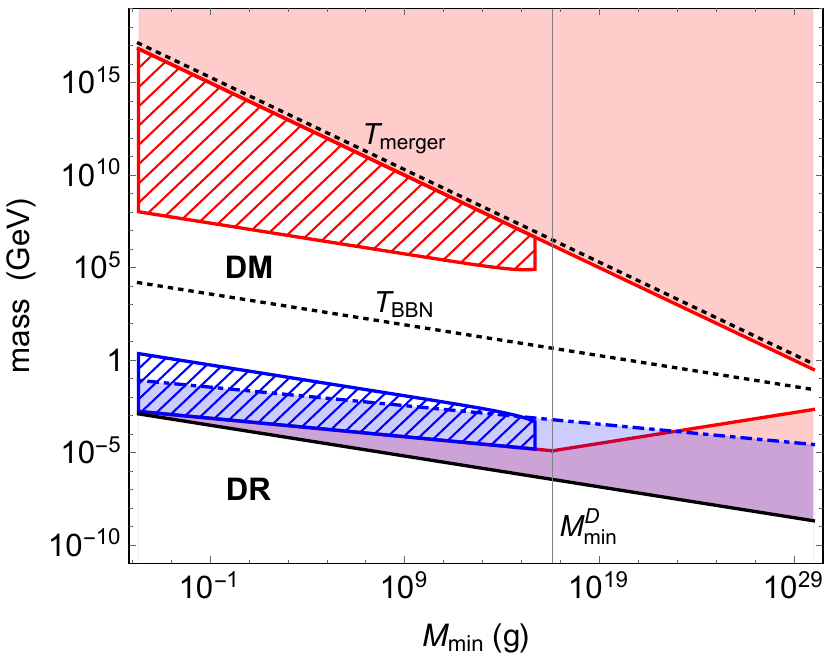}\label{fig:MH1}}\,\,
\subfloat[(b)\quad\quad With constrained $f$]{\includegraphics[width=7.6cm]{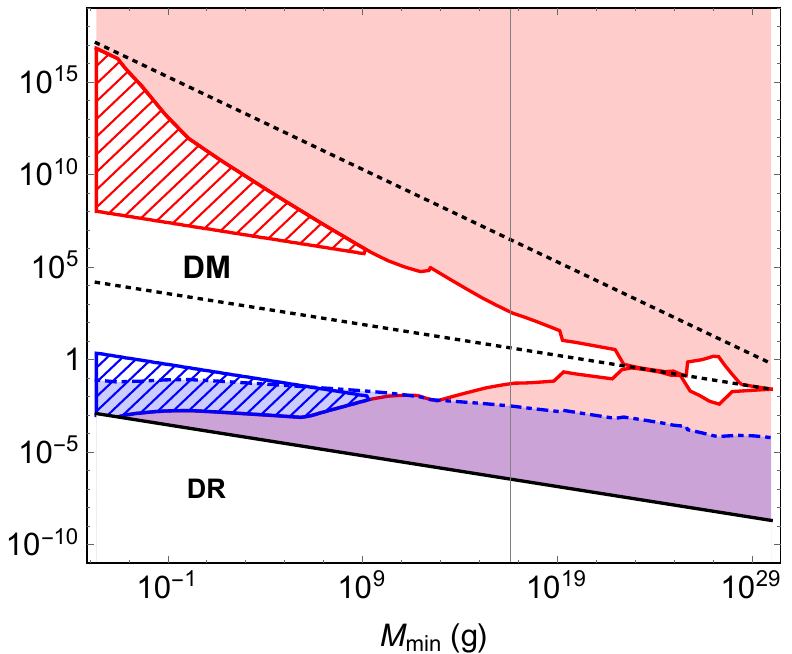}\label{fig:MH2}}
\end{tabular}
\caption{\label{MH} 
The allowed mass range of dark sector particles as a function of $\Mmin$, for the branching fraction $\textrm{B}_\chi=0.01-0.5$. 
In \ref{fig:MH1} the remnant abundance $f$ is taken as a free parameter and $f\leq 1/2$, whereas in \ref{fig:MH2} observational constrains in Fig.~\ref{fcons} are taken into account. 
The two white areas (including the red hatched region on white background) shows the allowed parameter space for dark matter (DM) and dark radiation (DR) respectively. The red boundary lines for the upper white area denote the abundance constraints for dark matter, and the hatched regions display the allowed parameter space in the domination case in particular, for $\Mmin\lesssim M_\textrm{min}^D$ given in (\ref{MD}). The blue region (including the blue hatched region) is excluded by the free-streaming constraints. $T_{\mathrm{BBN}}$, given in (\ref{eq:MBBN}), sets the lowest initial temperature. $T_{\mathrm{merger}}$, given in (\ref{eq:Tmerger}), denotes the temperature for the merger product of a remnant binary. 
}
\end{figure}

For the dark sector production, we have considered the requirement from the observed abundance and the free-streaming constraints for dark matter, and the contribution to the effective number of relativistic degrees of freedom, as parameterized by $\Delta N_\textrm{eff}$, for dark radiation.

Figure~\ref{MH} summarizes the allowed mass range for dark matter and dark radiation as functions of the remnant mass $\Mmin$. The upper white area (including the hatched region on white background) is for dark matter, where the dotted line in the middle denotes the lowest initial temperature $T_\mathrm{BBN}$ for the primordial 2-2-holes. The relic abundance cannot reach the observed value above or below the red solid boundary lines. For heavy dark matter case it is due to the limited number of particles produced from evaporation, while  for the light dark matter case it is due to a too small mass. The hatched areas show the relevant parameter space for the 2-2-hole domination scenario when $\Mmin\lesssim \Mmin^\textrm{D}$. The additional inner boundaries are related to the lower bound on the remnant abundance.  
The light dark matter case is subject to the free-streaming constraints in addition. As a result, the 2-2-hole domination is completely excluded, while a large range of parameter space remains viable for the non-domination case.   

For the strong coupling scenario where $f$ is not much constrained, the allowed mass range for dark matter is quite similar to that for the PBH production. Since we also considered the non-domination case, the dark matter mass can reach much lower scales than those considering only the domination case~\cite{Hooper:2019gtx}. In the weak coupling scenario, with increasing $\Mmin$, we see the relevant mass scale getting small and the allowed range shrinking. This is due to the decreasing initial temperature of 2-2-holes and the stronger constraints on the remnant abundance. We can see that the viable mass range remains large for the most strongly constrained $f$ cases, say $\Mmin\sim 10^{10}\,$g, indicating that a small fraction of 2-2-holes can still play a significant role for our understanding of dark matter. Heavy remnant cases with $\Mmin\gtrsim 10^{22}\,$g are excluded, except for a small window around $\Mmin\sim 10^{28}\,$ if a large dark sector is assumed. It points to a quite restricted mass range of dark matter that is slightly below GeV scale. 

For dark radiation, the mass upper bound ranges from $1\,$MeV for the strong coupling scenario to $1\,$eV for the heavy remnant case with $\Mmin\sim 10^{28}\,$g. The latter in particular is comparable to the case that dark radiation originates as a thermal relic. For the contribution to $\Delta N_\textrm{eff}$, as shown in Fig.~\ref{fig:DeltaN}, the 2-2-hole domination has the same prediction as in the case of black hole evaporation, which is mainly sensitive to the number of the degrees of freedom. The current limit requires $g_\textrm{DR}\lesssim 15$, while the future observations could probe down to $g_\textrm{DR}\approx 1$. The contribution in the non-domination case drops sharply for smaller $\Mini$. In the case of a large dark sector with $g_\textrm{DR}\gg 10$, the non-domination case could be relevant, and a nonzero $\Delta N_\textrm{eff}$ may point to a small mass range for $\Mini$. 

Finally, in contrast to black holes, dark sector particles can be reproduced at present by evaporation of the merger products of remnant binaries. Since the merger product acquires a very high temperature $T_{\mathrm{merger}}$, heavy particles produced with a suppressed rate before could be numerously produced now. From Fig.~\ref{MH}, we can see that $T_\textrm{merger}$ is in general much larger than the relevant mass scales and the emitted particles must be ultra-relativistic.
This then provides a natural realization of the boosted dark matter scenario, with the boost factor easily exceeding a few hundreds.
If dark sector particles only interact with SM gravitationally, direct detection could be challenging but still possible. For instance, a recent proposal considers an array of quantum-limited mechanical impulse sensors and demonstrates the capability of detecting the Planck-scale dark matter by using a large number of sensors~\cite{Carney:2019pza}. 

For the lower mass range, additional interaction with the SM may be required for the direct detection. If dark sector particles interact with hadrons through some mediators, the IceCube detectors could be the optimal targets for the highly boosted flux~\cite{Kopp:2015bfa}. Through deep inelastic scatterings, these energetic particles will create shower-like events as for the neutral current scattering of neutrinos~\cite{Ahlers:2018mkf}. 
Previously, the 2-2-hole remnant fraction was found to be mostly constrained by the measurements of photon and neutrino fluxes produced by the high energy emission of the remnant mergers~\cite{Aydemir:2020xfd}. But if the dark sector particles have a large number of degrees of freedom, the dark matter flux might provide the smoking gun signal for this process as long as its scattering cross section with hadrons is not too much smaller than that for neutrinos. For such cases, the dark matter relic abundance may receive additional contribution from the thermal production through freeze-out. It is possible to construct a dark sector model that predicts a subdominant thermal contribution due to a larger annihilation cross section, while being consistent with the current experimental constraints from the collider search and the direct detection. For instance, given what we know about the well-studied Higgs-portal or Z-portal dark matter models~\cite{Balazs:2017ple}, the constraints can be avoided if the dark matter mass is well above the TeV scale and the mediator mass is lighter but still considerably higher than the electroweak scale.\footnote{With a mediator heavier than electroweak scale but lighter than the dark matter mass, the direct detection constraints can be relaxed without much change on the relic abundance calculation.} As indicated in Fig.~\ref{MH}, there is still a large viable parameter space that may do the job for a wide range of $\Mmin$.    
We leave more detailed studies of particle physics models and the non-SM interactions of dark sector particles for future work.

\subsection{Baryogenesis}

\begin{figure}[h!]
\captionsetup[subfigure]{labelformat=empty}
\centering
\hspace{-0.8cm}
\begin{tabular}{lll}
\subfloat[(a)\quad  With unconstrained $f$]{\includegraphics[width=7.8cm]{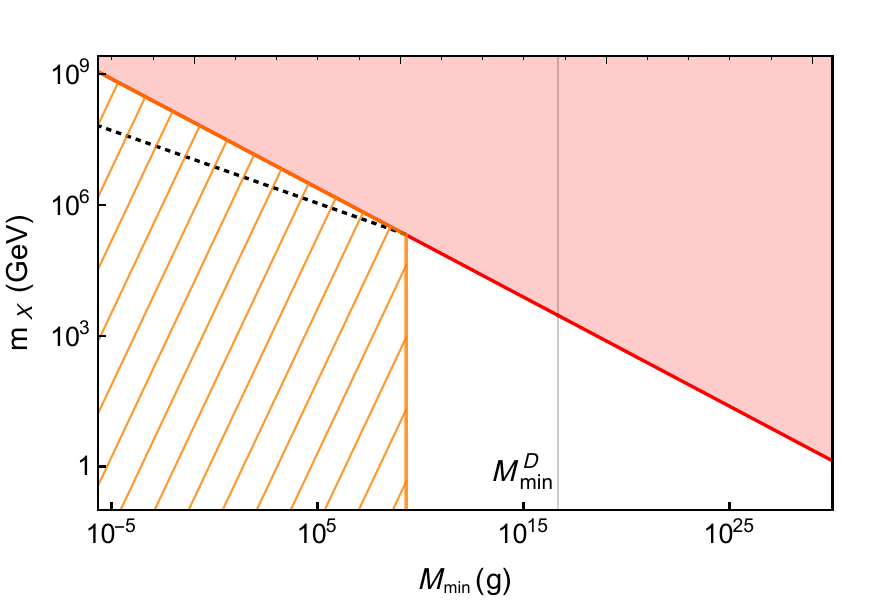}\label{fig:BaryoSuma}}
\subfloat[(b)\quad  With constrained $f$]{\includegraphics[width=7.8cm]{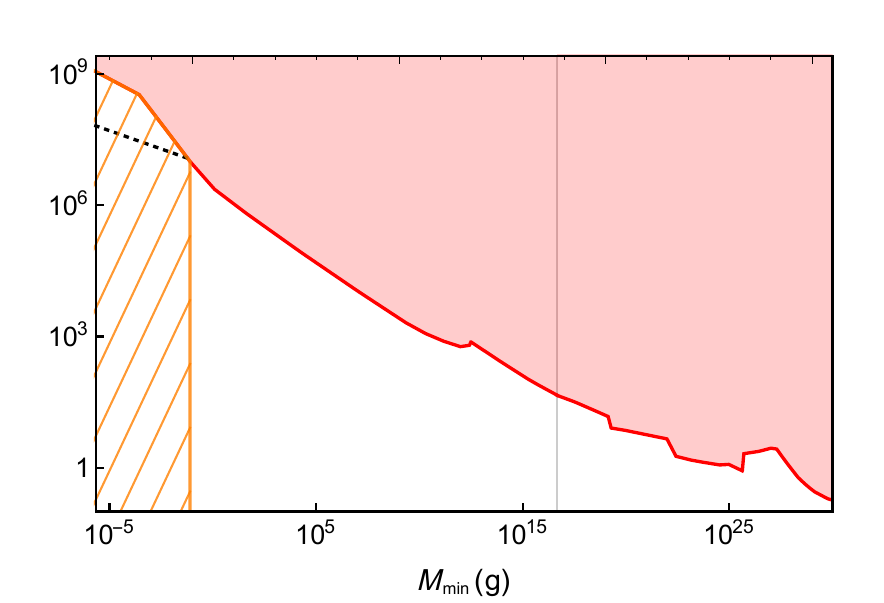}\label{fig:BaryoSumb}}
\end{tabular}
\caption{\label{BaryoSum} 
The upper bound on the decay particle mass for successful baryogenesis as a function of $\Mmin$, assuming the CP-violating parameter $\gamma\leq 1$ and $\mathrm{B}_X\leq 0.5$. In \ref{fig:BaryoSuma} the remnant abundance $f$ is taken as a free parameter and $f\leq 1$, whereas in \ref{fig:BaryoSumb} observational constrains in Fig.~\ref{fcons} are taken into account. The hatched region denotes the parameter space that allows leptogenesis, with the boundary value of $\Mmin$ defined by the intersection of the upper bound and the black dotted line, which shows the initial temperature for $\Mini=M_\textrm{EW}$, defined in (\ref{eq:MEW}).} 
\end{figure}

For the baryon asymmetry production, we have considered the out-of-equilibrium decay of heavy particles and electroweak baryogenesis. In the latter case, the total asymmetry produced by 2-2-holes scales with an inverse power of $\Mmin$, and even in the case with the Planck mass, a considerably large CP-violating phase is required to generate the observed value. As for the production through particle decays, our discussion applies to both baryogenesis through direct B-violating decays  and leptogenesis, depending on whether the background temperature after evaporation is smaller or larger than the electroweak scale. 

Figure~\ref{BaryoSum} shows the upper bound on the decay particle mass as a function of $\Mmin$ from the requirement of baryon asymmetry generation. As in the case of dark matter production, the particle has to be lighter for increasing $\Mmin$. 
Baryogenesis through direct B-violating decays can operate for a wide range of $\Mmin$, while leptogenesis is only allowed in the small region around $\Mmin=\Mp$ since the initial mass has to take a smaller value. 
For the former case,  the mass has to be smaller than  $\sim10^{9}$ GeV, and this is much lower than the expected range in GUT framework. 
For leptogenesis, although the right hand neutrino can stay light, the parameter space we show assumes a quite large CP-violating parameter and a large number of degrees of freedom for the decay particle. For a more realistic case, the available parameter space may disappear.

Evidently the parameter space we obtain requires caution for the model building. 
For direct baryogenesis, if the baryon number violation is provided through particles that lead to proton decay, one should be careful regarding the corresponding constraints. For instance, color-triplets in GUT models that couple to leptons and quarks are typically considered above $10^{11}$\,GeV in order to avoid proton decay, and are much larger than our highest upper limit $m_X\sim10^{9}$\,GeV in Fig.~\ref{BaryoSum}. This is generally enforced by the assumption that the triplet Yukawa couplings to the first-generation fermions are similar in magnitude to the Higgs Yukawa couplings, given that the triplet and the doublet come from the same multiplet of the SU(5). Nonetheless, their large mass hierarchy makes this assumption less motivated. Indeed, many mechanisms have been proposed to suppress the triplet Yukawa couplings, ensuring the proton stability for lighter particles, e.g. the Yukawa couplings with a suppression factor about $10^{-7}$ enable color-triplets in SUSY-SU(5) to be as light as $10^4$\,GeV (see the discussion in Ref.~\cite{Hooper:2020otu} and the references therein). Going beyond the GUT scenario, we can easily think of particles with appropriate quantum numbers, and write down baryon number violating interactions in a simplified model, without inducing the proton decay~\cite{Gu:2011ff,Arnold:2012sd}. These models are subject to much weaker observational constraints, e.g. neutron electric dipole moment and neutron-antineutron oscillation, and can provide successful baryogenesis with lighter particles. For instance, a color scalar with quantum number $\bar{6}$ can be as light as $10^4$\,GeV if the corresponding coupling is around $10^{-3}$~\cite{,Arnold:2012sd}.

As for leptogenesis, our constraint $m_X\lesssim10^{9}$\,GeV for the right-handed neutrino is much less challenging. In the simplest leptogenesis models for thermal production, where the lepton asymmetry is produced mainly by decay of the lightest right-handed neutrino, such low $m_X$ cannot satisfy the out-of-equilibrium condition~\cite{Davidson:2002qv}, and so the contribution from thermal production is  negligible. However, our parameter space for leptogenesis is more constrained than that for direct baryogenesis, and a relatively large CP-violation is required. It turns out that the "resonant leptogenesis" scenario is more relevant, where two right-handed neutrinos are nearly degenerate and then a great enhancement of CP-asymmetry can be produced. Given the decay width suppression, the maximum value for the CP-asymmetry $\gamma$ is around one for $m_X$ around TeV scale. This has a large overlap with our viable parameter space, and the relative importance of thermal production and 2-2-hole evaporation in this scenario deserves further study.    

Another question would be the implications of this limit in terms of the possible role of right-handed neutrinos in the seesaw mechanism. If one sets out to explain the smallness of the SM neutrino masses with $m_X\lesssim 10^9$\,GeV, the Yukawa coupling has to be smaller than $10^{-4}$, which may run into difficulties in a GUT framework if the Yukawa unification is required.
On the other hand, if the UV completion of the SM only includes the three right-handed neutrinos, a model mainly motivated by hierarchy arguments and the Higgs mass stability requirement suggests an upper bound $\sim10^6$\,GeV for the right-handed neutrino mass~\cite{Vissani:1997ys,Boyarsky:2009ix}. This is  not too far from our bound.

\section{Summary}
\label{sec:summary}
As a generic family of classical solutions in quadratic gravity, the 2-2-hole provides a probable endpoint of gravitational collapse as  an alternative to black holes. Since they are ultracompact and can be supermassive, 2-2-holes remain consistent with the current observations identified with black holes. Moreover, these objects do not possess event horizons due to the crucial role played by the quadratic curvature terms, and so they are free from the information-loss issue to begin with. A typical thermal 2-2-hole radiates like a black hole with the similar peculiar thermodynamic characteristics.  Primordial 2-2-holes could then evaporate at early universe and produce particles of all kinds. The evaporation leaves behind a 2-2-hole remnant, whose mass $\Mmin$ is determined from the mass of the additional spin-2 mode in the theory. Thus, any information on $\Mmin$ can help infer the new mass scale in quantum gravity. In a previous work~\cite{Aydemir:2020xfd}, we considered 2-2-hole remnants as all dark matter and the constraints from various observations. The parameter space is considerably restricted, favoring toward the Planck mass remnants, namely,  the strong coupling scenario for quadratic gravity.

In this work, we have continued our phenomenological investigation for primordial 2-2-holes. By abandoning the condition of remnants as all dark matter, we could consider remnants much heavier than the Planck mass. We have investigated the scenario that the majority of dark matter consists of particles produced by early time evaporation, while the remnant contribution accounts for the rest. We have also considered the possible dark radiation contribution to $\Delta N_\textrm{eff}$ and explored different mechanisms for baryon asymmetry generation from 2-2-hole evaporation. The implications can be quite different if there is an era of 2-2-hole domination in the early universe, which can be realized only for $\Mmin\lesssim 10^{16}$\,g.  Throughout the paper, both domination and non-domination scenarios have been considered, when necessary. 

We have found that the primordial 2-2-hole picture can accommodate both dark matter production and  baryogenesis through decay of heavy particles for a large range of $\Mmin$, including the heavier remnant cases subject to strong abundance constrains. In the weak coupling scenario, the relevant particle mass scales get smaller with increasing $\Mmin$ due to the lower value of the initial temperature. The parameter space is less restricted for smaller $\Mmin$, which in the Planck mass limit converges to the case of PBHs with the Planck mass remnants. Considering the abundance requirement and the free-streaming constraints, the dark matter mass can vary from $10^{17}\,$GeV to $0.1\,$GeV for $\Mmin\sim\Mp$--$10^{28}$\,g. 
As for baryon asymmetry generation, baryogenesis through direct B-violating decays can operate for a wide range of $\Mmin$ values, while leptogenesis is only allowed within a small window close to the Planck mass. The upper mass limit for the decay particle is $10^{9}$\,GeV due to the existence of remnants, and this requires caution in model building for direct B-violating decays.  
For dark radiation contribution to $\Delta N_\textrm{eff}$, the domination scenario makes a simple prediction depending only on the number of degrees of freedom, and the current data requires it to be smaller than $15$. The contribution in the non-domination case is in general significantly suppressed, but it may be relevant for a small window of the 2-2-hole initial mass if there is a large dark sector. 
In contrast to black holes, 2-2-hole remnants can reproduce these particles at present through strong radiation from the merger products. This may provide additional means for the test of the production mechanisms discussed in this paper.


\vspace{0.1cm}
\section*{Acknowledgements} 
\vspace{-0.1cm}
Work of U.A. is supported in part by the Chinese Academy of Sciences President's International Fellowship Initiative (PIFI) under Grant No. 2020PM0019, and the Institute of High Energy Physics, Chinese Academy of Sciences, under Contract No.~Y9291120K2. J.R. is supported by the Institute of High Energy Physics under Contract No. Y9291120K2.


\appendix

\section{More on free-streaming constraints}
\label{sec:appdA}

In this appendix, we consider another derivation for the free-streaming constraints. It involves the momentum distribution function of dark matter and constraints on the fraction of relativistic particles~\cite{Lennon:2017tqq}. The final results agree with the simple estimation in Sec.~\ref{sec:DM} up to an order one factor. 

After the 2-2-hole evaporation, the particle spectrum is a superposition of the earlier time  instantaneous emissions with the corresponding redshift, 
\begin{eqnarray}
F(p,t)=\int_{t_\chi}^{\tau_L}d\tau \frac{d\dot{N}}{dp}\left(p \frac{a(t)}{a(\tau)},\tau\right)\frac{a(t)}{a(\tau)}\,,
\end{eqnarray}
where $t_\chi$ is defined in (\ref{eq:tchiLH}). The instantaneous emissions follow the Planck distribution,
\begin{eqnarray}
 \frac{d\dot{N}}{dp}(p,t)\approx\frac{2M^2(t)}{\pi\, \Mp^4}\frac{p^2}{e^{p/T(t)}-1}\,,
\end{eqnarray}
with $M(t),\,T(t)$ in (\ref{eq:LMlimitTime0}). Since the distribution at $t>\tau_L$ is simply a redshift of $F(p,\tau_L)$, an upper bound $f_\textrm{S}$ on the fraction of relativistic particles at some later time $t_\textrm{S}$ can be translated as a constraint for $F(p,\tau_L)$, with  
\begin{eqnarray}\label{eq:freestreaming2}
\frac{\int_{p_\textrm{min}}^\infty dp \,F(p,\tau_L)}{\int_0^\infty dp \, F(p,\tau_L)}<f_\textrm{S},\quad
p_{\textrm{min}}=m_{\chi}\frac{a(t_\textrm{S})}{a(\tau_L)}\,.
\end{eqnarray}

\begin{figure}[h]
  \centering%
{ \includegraphics[width=12cm]{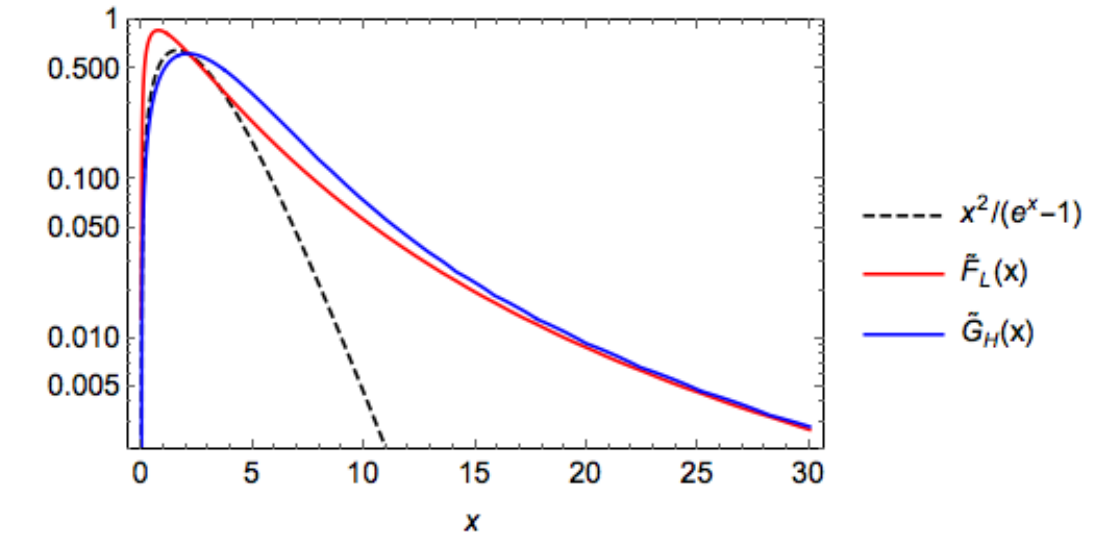}}
\caption{\label{pfunc} 
The dimensionless momentum distribution functions for light dark matter (red) and heavy dark matter (blue). For comparison, the dot dashed line shows the Planck distribution.}
\end{figure}

For the light mass case, with the starting time $t_\chi=t_\textrm{init}$, we find the momentum distribution function
\begin{eqnarray}
F_L(p,\tau_L)\approx \frac{2\, \tau_L}{\pi\, \Mp^4} \Mini^2\, T_\textrm{init}^2\, \tilde{F}_L(p/T_\textrm{init})\,,
\end{eqnarray}
where the momentum dependence is fully encoded in the dimensionless function
\begin{eqnarray}
\tilde{F}_L(a)\approx a^2 \int_0^1 dx\,\frac{(1-x)^{2/3}}{x^{3/2}} \left[e^{a\frac{(1-x)^{1/3}}{x^{1/2}}}-1\right]^{-1}\,.
\end{eqnarray}
As shown in Fig.~\ref{pfunc}, $\tilde{F}_L(p/T_\textrm{init})$ has a long tail in comparison to the Planck distribution. Assuming $f_\textrm{S}\approx 10\%$, $t_\textrm{S}\approx 10^6\,$s and one dominant component of light dark matter, the free-streaming constraint (\ref{eq:freestreaming2}) implies $m_\chi\, a(t_S)/(T_\textrm{init} a(\tau_L))\gtrsim 10$, i.e. 
\begin{eqnarray}\label{eq:FSLD2}
m_\chi\gtrsim 3.1\times 10^{-5}\,\hMmin^{-1/2} \, \hMini^{1/2}\,\textrm{GeV}\,.
\end{eqnarray} 
It takes the same form as the simpler estimation (\ref{eq:FSLD1}), and the coefficient is numerically close. 

For the heavy mass case, we find the momentum distribution function
\begin{eqnarray}
F_H(p,\tau_L)\approx \frac{2\, \tau_L}{\pi\, \Mp^4} \Mini^2\, T_\textrm{init}^2 \,\tilde{F}_H\left(\frac{p}{T_\textrm{init}}, \frac{m_\chi}{T_\textrm{init}}\right)\,,
\end{eqnarray}
where the form factor depends on $m_\chi$ in addition, 
\begin{eqnarray}
\tilde{F}_H(a,b)\approx a^2 \int_{1-b^{-3}}^1 dx \,\frac{(1-x)^{2/3}}{x^{3/2}}\left[e^{a\frac{(1-x)^{1/3}}{x^{1/2}}}-1\right]^{-1}
\approx b^{-3} \tilde{G}_H(a/b)\,.
\end{eqnarray}
The momentum dependence now comes solely from the dimensionless function $\tilde{G}_H(p/m_\chi)$ and is independent of $T_\textrm{init}$. As shown in Fig.~\ref{pfunc}, $\tilde{G}_H(x)$ approaches $\tilde{F}_L(x)$ at large $x$.  The free-streaming constraint for one dominant component case is then $a(t_S)/a(\tau_L)\gtrsim 10$. As in the simple estimation, this condition is independent of $m_\chi$ and can be easily satisfied for $\Mini\lesssim M_\textrm{BBN}$.

\raggedright  
\bibliography{References_H0}{}
\bibliographystyle{apsrev4-1}

\end{document}